# QoS-Aware Base-Station Selections for Distributed MIMO Links in Broadband Wireless Networks

Qinghe Du, *Student Member, IEEE,* and Xi Zhang, *Senior Member, IEEE*

***Abstract*—The distributed multiple-input-multiple-output (MIMO) techniques across multiple cooperative base-stations (BS) can significantly enhance the capability of the broadband wireless networks in terms of quality-of-service (QoS) provisioning for wireless data transmissions. However, the computational complexity and the interfering range of the distributed MIMO systems also increase rapidly as the number of cooperative BS's increases. In this paper, we propose the QoS-aware BS-selection schemes for the *distributed wireless MIMO links*, which aim at minimizing the BS usages and reducing the interfering range, while satisfying diverse statistical delay-QoS constraints characterized by the delay-bound violation probability and the effective capacity technique. In particular, based on the channel state information (CSI) and QoS requirements, a subset of BS with variable cardinality for the distributed MIMO transmission is dynamically selected, where the selections are controlled by a central server. For the *single-user* scenario, we develop two optimization frameworks, respectively, to derive the efficient BS-selection schemes and the corresponding resource allocation algorithms. One framework uses the incremental BS-selection and time-sharing (IBS-TS) strategies, and the other employs the ordered-gain based BS-selection and probabilistic transmissions (OGBS-PT). The IBS-TS framework can yield better performance, while the scheme developed under the OGBS-PT framework is easier to implement. For the *multi-user* scenario, we propose the optimization framework applying the priority BS-selection, block-diagonalization precoding, and probabilistic transmission (PBS-BD-PT) techniques. We also propose the optimization framework applying the priority BS-selection, time-division-multiple-access, and probabilistic transmission (PBS-TDMA-PT) techniques. We derive the optimal transmission schemes for all the aforementioned frameworks, respectively. Also conducted is a set of simulation evaluations which compare our proposed schemes with several baseline schemes and show the impact of the delay-QoS requirements, transmit power, and traffic loads on the performances of BS selections for distributed MIMO systems.**

## I. Introduction

In order to increase the coverage of broadband wireless networks, distributed multiple-input-multiple-output (MIMO) techniques, where multiple location-independent base stations (BS) cooperatively transmit data to mobile users, have attracted more and more research attentions [1]–[4]. In particular, the distributed MIMO techniques can effectively organize multiple location-independent BS's to form the distributed MIMO links connecting with mobile users, while not requiring too many multi-antennas, which are expensive, equipped at individual BS's. Like the conventional centralized MIMO system [5]–[7], the distributed MIMO system can significantly enhance the capability of the broadband wireless networks in terms of the quality-of-service (QoS) provisioning for wireless transmissions as compared to the single antenna system. However, the distributed nature for cooperative multi-BS transmissions also imposes many new challenges in wide-band wireless communications, which are not encountered in the centralized MIMO systems.

First, the cooperative distributed transmissions cause the severe difficulty for synchronization among multiple location-independent BS transmitters. Second, as the number of cooperative BS's increases, the computational complexity for MIMO signal processing and coding also grow rapidly. Third, because the coordinated BS's are located at different geographical positions, the cooperative communications in fact enlarge the interfering areas for the used spectrum, thus drastically degrading the frequency-reuse efficiency in the spatial domain. Finally, many wide-band transmissions are sensitive to the delay, and thus we need to design QoS-aware distributed MIMO techniques, such that the scarce wireless resources can be more efficiently utilized.

Towards the above issues, many research works on distributed MIMO transmissions have been proposed recently. The feasibility of transmit beamforming with efficient synchronization techniques over distributed MIMO link has been demonstrated through experimental tests and theoretical analyses [2], [3], suggesting that complicated MIMO signal processing techniques are promising to implement in realistic systems. While the antenna selection [6], [7] is an effective approach to reduce the complexity for centralized MIMO systems, which can be also extended to distributed MIMO systems for the BS selection. It is clear that the BS-selection techniques can significantly decrease the processing complexity, while still achieving high throughput gain over the single BS transmission. Also, it is desirable to minimize the number of selected BS's through BS-selection techniques, which can effectively decrease the interfering range and thus improve the frequency-reuse efficiency of the entire wireless network. Most previous research works for BS/antenna selections mainly focused on the scenarios of selecting a subset of BS's/antennas with the fixed cardinality [4], [6], [7]. However, it is evident that based on the wireless-channel status, BS-subset selections with dynamically adjusted cardinality can further decrease the BS usage. More importantly, how to efficiently support diverse

The research reported in this paper was supported in part by the U.S. National Science Foundation CAREER Award under Grant ECS-0348694.

The authors are with the Networking and Information Systems Laboratory, Department of Electrical and Computer Engineering, Texas A&M University, College Station, TX 77843 USA (e-mails: duqinghe@tamu.edu; xizhang@ece.tamu.edu).

delay-QoS requirements through BS-selection in distributed MIMO systems still remains as a widely cited open problem.

To overcome the aforementioned problems, we propose the QoS-aware BS-selection schemes for the distributed wireless MIMO links, which aim at minimizing the BS usages and reducing the interfering range, while satisfying diverse statistical delay-QoS constraints. In particular, based on the channel state information (CSI) and QoS requirements, the subset of BS with variable cardinality for the distributed MIMO transmission is dynamically selected, where the selections are controlled by a central server. For the *single-user* scenario, we consider the optimization framework which uses the incremental BS-selection and time-sharing (IBS-TS) strategies, and study another framework which employs the ordered-gain based BS-selection and probabilistic-transmissions (OGBS-PT) techniques. For the *multi-user* scenario, we propose the optimization framework applying the priority BS-selection, block-diagonalization multiple-access, and probabilistic transmission (PBS-BD-PT) techniques. We also propose the optimization framework applying the priority BS-selection, time-division-multiple-access, and probabilistic transmission (PBS-TDMA-PT) techniques. We derive the optimal transmission schemes for the above frameworks, respectively, and conduct comparative analyses with the baseline schemes through simulations.

The rest of this paper is organized as follows. Section II describes the system model for distributed MIMO transmissions. Section III introduces the statistical QoS guarantees and the concept of effective capacity. Section IV proposes the optimization framework for QoS-aware BS sections of the single-user case and develops its corresponding optimal solution. Section V develops the optimization framework for multi-user case and derives its optimal solution. Section VI simulates our proposed schemes. The paper concludes with Section VII.

## II. System Model

### A. System Architecture

We concentrate on the wireless *distributed MIMO* system for downlink transmissions depicted in Fig. 1, which consists of $K_{\rm bs}$ distributed BS's, $K_{\rm mu}$ mobile users, and one central server. The $m$th BS has $M_m$ transmit antennas for $m = 1, 2, \ldots, K_{\rm bs}$ and the $n$th mobile user has $N_n$ receive antennas for $n = 1, 2, \ldots, K_{\rm mu}$. All distributed BS's are connected to the central server through high-speed optical connections. The data to be delivered to the $n$th mobile user, $n = 1, 2, \ldots, K_{\rm mu}$, arrives at the central server with a constant rate, which is denoted by $\overline{C}_n$. Then, the central server dynamically controls these distributed BS's to cooperatively transmit data to the mobile users under the specified delay-QoS requirements.

For the case of $K_{\rm mu} = 1$, the distributed BS's and the mobile user form a single wireless MIMO link; when $K_{\rm mu} \geq 2$, the distributed BS's and the mobile users form the broadcast MIMO link for data transmissions. The wireless fading channels between the $m$th BS and the $n$th mobile user is modeled by an $N_n \times M_m$ matrix $\mathbf{H}_{n,m}$. The element at the $i$th row and $j$th column of $\mathbf{H}_{n,m}$, denoted by $(\mathbf{H}_{n,m})_{i,j}$, is the complex channel gain between the $i$th receive antenna of $n$th mobile user and the $j$th transmit antenna of the $m$th BS. All elements of $\mathbf{H}_{n,m}$ are independent and circularly symmetric complex Gaussian random variables with zero mean and the variance equal to $\overline{h}_{n,m}$. Also, the instantaneous aggregate power gain of the MIMO link between the $n$th mobile user and the $m$th BS, denoted by $\gamma_{n,m}$, is defined by

$$\gamma_{n,m} \triangleq \frac{1}{M_m} \sum_{i=1}^{N_n} \sum_{j=1}^{M_m} \left| (\mathbf{H}_{n,m})_{i,j} \right|^2. \tag{1}$$

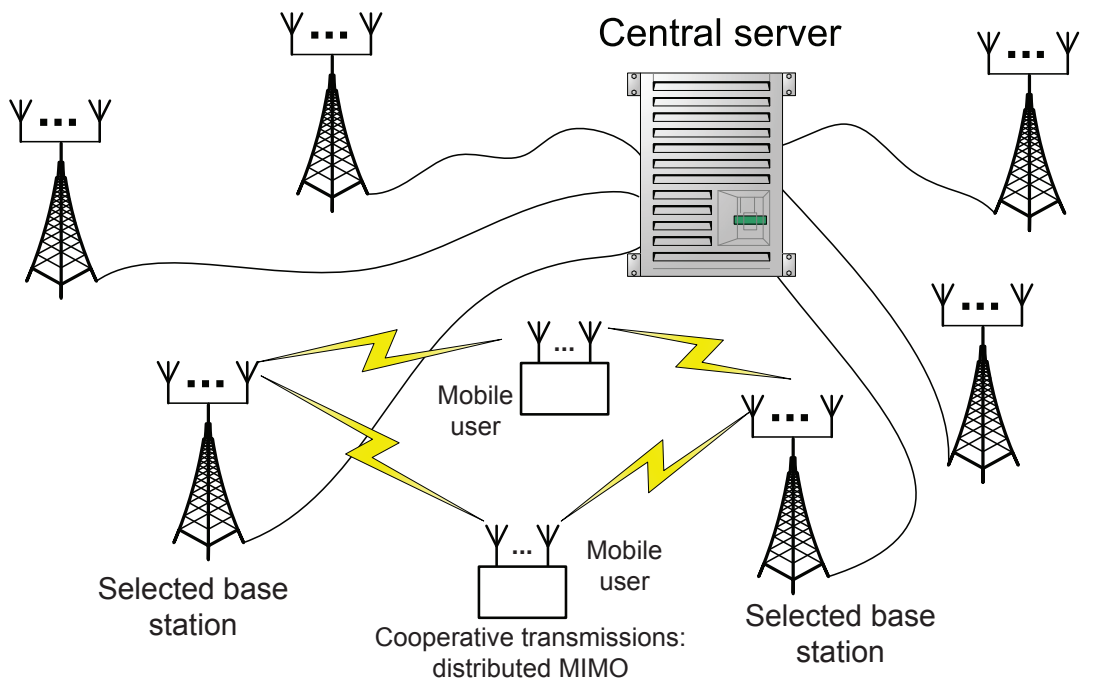

Fig. 1. System model of a wireless distributed MIMO system for downlink transmissions.

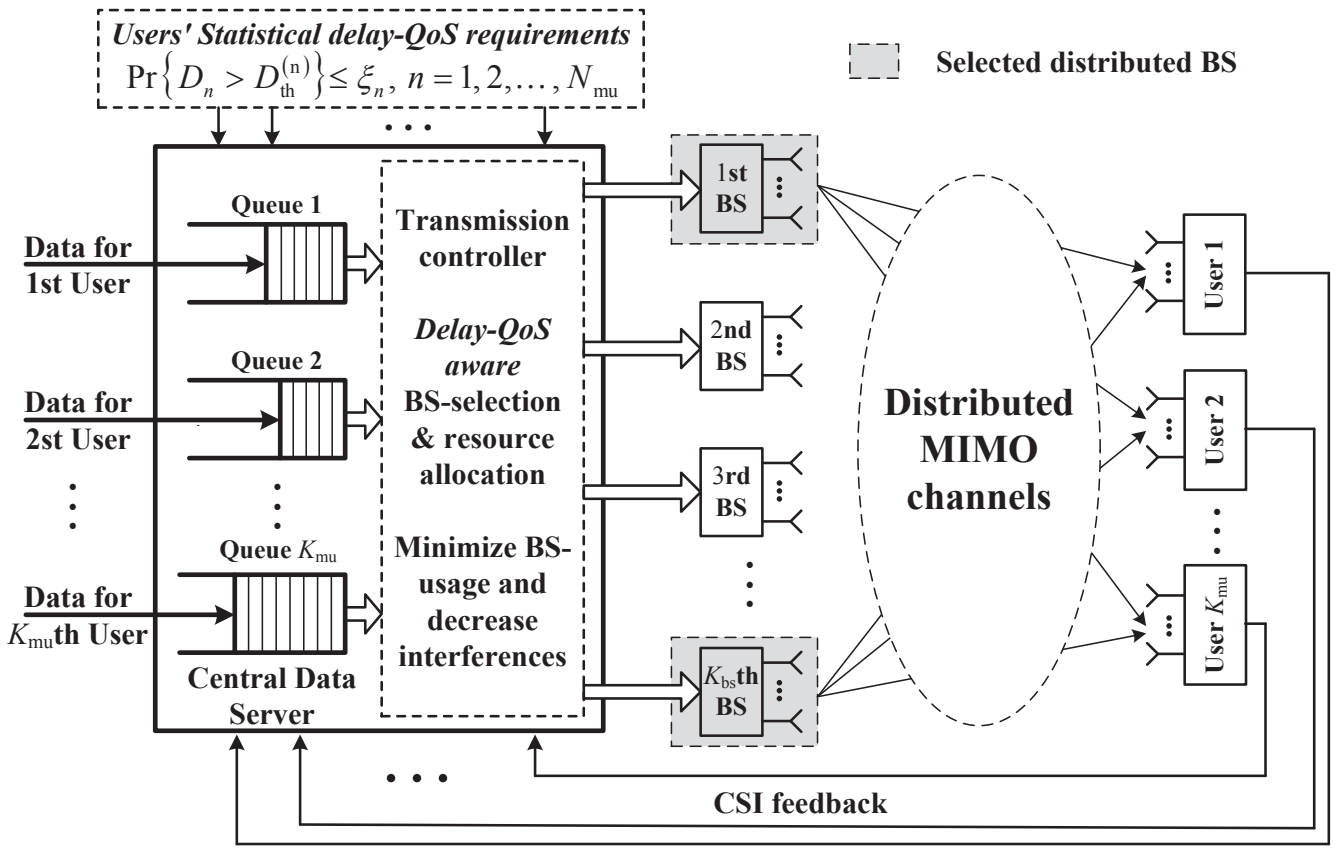

Fig. 2. Our proposed QoS-aware BS-selection framework.

We define $\mathbf{H}_n \triangleq [\mathbf{H}_{n,1} \ \mathbf{H}_{n,2} \ \cdots \ \mathbf{H}_{n,K_{\rm bs}}]$ as the CSI for the $n$th mobile user for $n = 1, 2, \ldots, K_{\rm mu}$. The matrix $\mathbf{H}_n$ follows the independent block-fading model, where $\mathbf{H}_n$ does not change within a time period with the fixed length $T$, called a time frame, but varies independently from one frame to the other frame. Furthermore, we define $\mathbf{H} \triangleq [\mathbf{H}_1^{\tau} \ \mathbf{H}_2^{\tau} \ \cdots \ \mathbf{H}_{K_{\rm bs}}^{\tau}]^{\tau}$, representing a fading state of the entire distributed MIMO system, where the superscript $\tau$ denotes the transpose operation on a matrix or a vector.

Under the aforementioned model, our proposed BS-selection framework is illustrated in Fig. 2. As shown in Fig. 2, based on CSI feedback from the users, the central data server will dynamically select a subset of BS's and then use the transmit antennas of all selected BS's to construct the distributed channels to transmit data to the $K_{\rm mu}$ users. Our BS-selection strategies and the corresponding resource allocation algorithms depends not only on the CSI, but also the *statistical delay-QoS constraints* (to be detailed on Section II-B and Section III) for the incoming traffics. Although the BS/antenna selection techniques for distributed/centralized MIMO systems have been extensively studied, most existing works [4], [6], [7] focused on maximizing the capacity or minimizing the error rate given the specified BS/antenna subset cardinality. In contrast, our work in this paper aims at tackling the following new challenges for distributed MIMO system. 1) Our BS-selection and the associated resource allocation algorithms need to guarantee the specified delay-QoS requirements for the incoming traffics. 2) The cardinality of selected BS-subset vary with CSI and delay QoS, which is more flexible and efficient than the selection schemes with the fixed subset cardinality. 3) We aim at using the minimum average number of BS's to support the incoming traffic with delay-QoS guarantees, which also decreases the interferences to the entire network.

### B. The Delay QoS Requirements

The central data server maintains a queue for the incoming traffic to each mobile user. We mainly focus on the queueing delay in this paper because the wireless channel is the major bottleneck for high-rate wireless transmissions. Since it is usually unrealistic to guarantee the hard delay bound over the highly time-varying wireless channels, we employ the statistical metric, namely, the *delay-bound violation probability*, to characterize the diverse delay QoS requirements. Specifically, for the $n$th mobile user, the probability of violating a specified delay bound, denoted by $D_{\rm th}^{(n)}$, cannot exceed a given threshold $\xi_n$. That is, the inequality

$$\Pr\left\{D_n > D_{\rm th}^{(n)}\right\} \le \xi_n, \quad n = 1, 2, \ldots, N_{\rm mu}, \tag{2}$$

needs to hold, where $D_n$ denotes the queueing delay in the $n$th mobile user's queueing system.

### C. Performance Metrics and Design Objective

We denote by $L$ the cardinality of the selected BS subset (the number of selected BS's) for the distributed MIMO transmission in a fading state. Then, we denote the expectation of $L$ by $\overline{L}$ and call it the *average BS usage*. As mentioned in Section II-A, our major objective is to minimize $\overline{L}$ through dynamic BS selection while guaranteeing the delay QoS constraint specified by Eq. (2). Besides the average BS usage, we also need to evaluate the *average interfering range* affected by the distributed MIMO transmission. The instantaneous interfering range, denoted by $A$, is defined as the area of the region where the average received power under the current MIMO transmission is larger than a certain threshold denoted by $\sigma_{\rm th}^2$. The average interfering area is then defined as the expectation $\mathbb{E}\{A\}$ over all fading states. Clearly, minimizing $\overline{L}$ can not only reduce implementation complexity, but also decrease the average interfering range affected by the transmit power.

### D. The Power Control Strategy

The transmit power of our distributed MIMO system varies with the number of selected BS's. In particular, given the number $L$ of selected BS's, the total instantaneous transmitted power used for distributed MIMO transmissions is set as a constant equal to $\mathcal{P}_L$. Furthermore, $\mathcal{P}_L$ linearly increases with $L$ by using the strategy as follows:

$$\mathcal{P}_L = \mathcal{P}_{\rm ref} + \kappa(L-1), \quad L = 1, 2, \ldots, K_{\rm bs}, \tag{3}$$

where $\mathcal{P}_{\rm ref} > 0$ is called the *reference power* and $\kappa \ge 0$ describes the power increasing rate against $L$. Also, we define $\mathcal{P}_L \triangleq 0$ for $L = 0$. The above power adaptation strategy is simple to implement, while the average transmit power can be effectively decreased through minimizing the average number of used BS's. In addition, Eq. (3) can upper-bound the instantaneous interferences and the interfering range over the entire network.

## III. Effective Capacity Approach for Statistical Delay-QoS Guarantees

In this paper, we apply the theory of statistical QoS Guarantees [8], [9], [13], [20] to integrate the constraint on delay-bound violation probability given by Eq. (2) into our BS selection design. Consider a stable dynamic discrete-time queueing system. The data arrival-rate and data service-rate of the queueing system are denoted by $C[k]$ and $R[k]$, respectively, where $k$ is the index for the time frame with a fixed time-duration equal to $T$ (as also described in Section II-A) and the units of $C[k]$ and $R[k]$ are nats/frame. The $C[k]$ and $R[k]$ change from frame to frame and thus can be characterized as the time-varying processes. By using the asymptotic analyses based on the large deviation principal, the author of [8] showed that under the sufficient conditions, the queue-length process as a function of $t$, denoted by $Q[k]$, converges in distribution to a random variable $Q$ satisfying the following equation:

$$-\lim_{Q\to\infty} \frac{\log(\Pr\{Q > Q_{\rm th}\})}{Q_{\rm th}} = \theta, \tag{4}$$

for a certain $\theta > 0$. A set of sufficient conditions given in [8] for Eq. (4) are summarized as follows. 1) Both $C[k]$ and $R[k]$ are stationary. 2) The Gartner-Ellis limits [8] for $\mathcal{C}[k]$ and $\mathcal{R}[k]$, denoted by $\Lambda_C(\theta)$ and $\Lambda_R(\theta)$, respectively, exist for all $\theta$, where $\mathcal{C}[k] \triangleq \sum_{t=1}^{k} C[t]$ and $\mathcal{R}[k] \triangleq \sum_{t=1}^{k} R[t]$. 3) The processes $C[k]$ and $R[k]$ are independent. 4) There exists a certain $\theta > 0$ such that $\Lambda_C(\theta) = \Lambda_R(-\theta)$. For the details of the sufficient conditions for Eq. (4), please refer to [8]. It has been shown that Eq. (4) holds for many stable queueing systems with typical arrival/departure processes [8], [13], such as Markovian processes, auto-regressive processes, and capacity-achieving service processes over i.i.d. block-fading wireless channels.

Based on Eq. (4), the probability that the queue-length, denoted by $Q$, exceeding a given bound $Q_{\rm th}$ can be approximated [8] by

$$\Pr\{Q > Q_{\rm th}\} \approx e^{-\theta Q_{\rm th}}, \tag{5}$$

where $\theta > 0$ is a constant called QoS exponent. It is clear that the larger (smaller) $\theta$ implies the lower (higher) queue-length-bound violation probability. Furthermore, when the delay bound becomes the main QoS metric of interests, the delay-bound violation probability can be approximated [9], [20] by

$$\Pr\{D > D_{\rm th}\} \approx e^{-\theta\varphi(\theta)D_{\rm th}}, \tag{6}$$

where $D$ and $D_{\rm th}$ denote the queueing delay and delay bound, respectively, and $\varphi(\theta)$ is known as the effective bandwidth [8] of the arrival-rate process under the given $\theta$. When the arrival rate $C[k]$ is equal to a constant $\overline{C}$ over all $k$ and the departure rate $R[k]$ is time-varying, Eq. (6) can be written [9], [20]–[22] as

$$\Pr\{D > D_{\rm th}\} \approx e^{-\theta\overline{C}D_{\rm th}}. \tag{7}$$

Then, to upper-bound $\Pr\{D > D_{\rm th}\}$ with a threshold $\xi$, using Eq. (7), we get the minimum required QoS exponent $\theta$ as follows:

$$\theta = -\frac{\log(\xi)}{\overline{C}D_{\rm th}}. \tag{8}$$

Consider a discrete-time arrival process with constant rate $\overline{C}$ and the discrete-time time-varying departure process $R[k]$, where $k$ is the time index. In order to guarantee the desired $\theta$ determined by Eq. (8), the statistical QoS theory [8]–[10] shows that the *effective capacity* $\mathcal{C}(\theta)$ of the service-rate process $R[k]$ needs to satisfy

$$\mathcal{C}(\theta) \geq \overline{C}, \tag{9}$$

under the given QoS exponent $\theta$. The *effective capacity* function is defined in [9] as the maximum constant arrival rate which can be supported by the service rate to guarantee the specified QoS exponent $\theta$. If the service-rate sequence $R[k]$ is stationary and time uncorrelated, the effective capacity can be written [13] as

$$\mathcal{C}(\theta) \triangleq -\frac{1}{\theta}\log\left(\mathbb{E}\left\{e^{-\theta R[k]}\right\}\right), \tag{10}$$

where $\mathbb{E}\{\cdot\}$ denotes the expectation.

In our distributed MIMO system, the BS selection result is designed as the function determined by the current CSI. Thus, the corresponding transmission rate (service rate) is time independent under the independent block-fading model (see Section II-A). Then, applying Eqs. (8)-(9), the delay QoS constraints given by Eq. (2) can be equivalently converted to:

$$\mathbb{E}_{\mathbf{H}}\left\{e^{-\theta_n R_n} - e^{-\theta_n\overline{C}_n}\right\} \leq 0, \quad ,n = 1,2,\ldots,N_{\rm mu}, \tag{11}$$

where $\theta_n = -\log(\xi_n)\Big/\left(\overline{C}_n D_{\rm th}^{(n)}\right)$ and $\mathbb{E}_{\mathbf{H}}\{\cdot\}$ denotes the expectation over all $\mathbf{H}$.

## IV. QoS-Aware BS Selection for the Single-User Case

We focus on the scenario with a single mobile user in this section, where $K_{\rm mu} = 1$. For presentation convenience, we use the term *transmission mode* $L$ to denote the case where the cardinality of the selected BS subset is equal to $L$. Given transmission mode $L$, we denote by $\Omega_L$ the set of indices of selected BS's, where $\Omega_L = \{i_{L,1}, i_{L,2}, \ldots, i_{L,L}\}$ and $i_{L,\ell} \in \{1,2,\ldots,K_{\rm bs}\}$ for $\ell = 1,2,\ldots,L$. Once a BS is selected, we use all its transmit antennas for data transmissions. Then, we characterize the channel matrix for the selected BS subset by $\mathbf{H}_{\Omega_L}$ and write $\mathbf{H}_{\Omega_L}$ as $\mathbf{H}_{\Omega_L} \triangleq \left[\mathbf{H}_{1,i_{L,1}}\ \mathbf{H}_{1,i_{L,2}}\ \cdots\ \mathbf{H}_{1,i_{L,L}}\right]$, which is an $N_1 \times \mathcal{M}_L$ matrix with $\mathcal{M}_L \triangleq \sum_{\ell=1}^{L} M_{i_{L,\ell}}$. Accordingly, the physical-layer signal transmission is characterized by

$$\mathbf{y} = \mathbf{H}_{\Omega_L}\mathbf{s}_{\Omega_L} + \boldsymbol{\varsigma},$$

where $\mathbf{y}$ is the $N_1 \times 1$ received signal vector and $\boldsymbol{\varsigma}$ denotes the $N_1 \times 1$ additive Gaussian noise vector whose elements are independent with unit power. The variable $\mathbf{s}_{\Omega_L} \triangleq \left[\mathbf{s}_{i_{L,1}}^{\tau}, \mathbf{s}_{i_{L,2}}^{\tau}, \ldots, \mathbf{s}_{i_{L,L}}^{\tau}\right]^{\tau}$ is the input signal vector (transmitted signal vector) for the MIMO channel $\mathbf{H}_{\Omega_L}$, where $\mathbf{s}_{i_{L,\ell}}$ is the $M_{i_{L,\ell}} \times 1$ signal vector transmitted from the $(M_{i_{L,\ell}})$-th BS.

Clearly, for dynamic BS selections of distributed MIMO transmissions, we need to answer the following three questions: (i) For a specified transmission mode $L$, how do we determine the BS subset $\Omega_L$? (ii) How are the wireless resources shared if applying multiple modes within a time frame? (iii) Which transmission modes will be used and how to quantitatively allocate the wireless resources? We first study associated issues for question (i) in Section IV-A. Then, we introduce the time-sharing transmission and probabilistic transmission to share the resources across different modes in Section IV-B. Following the discussions in Sections IV-A and IV-B, we formulate two analytical optimization frameworks to answer question (iii). In particular, the first optimization framework is based on the incremental BS-selection algorithm and the time-sharing strategy; the second framework applies the ordered-gain based BS selection algorithm with probabilistic transmission, which will be detailed in Sections IV-C and IV-D, respectively. The BS-selection scheme derived under the first optimization framework can achieve better performance, while the second optimization framework is simpler to implement.

### *A. BS-Selection Strategy Given the Cardinality of the BS Subset*

In this paper, we focus on the spatial multiplexing based MIMO transmissions. Given $\Omega_L$ in a fading state, the maximum achievable data rate (Shannon capacity), denoted by $R(\Omega_L)$ (nats/frame), is determined [5] by

$$R(\Omega_L) = \max_{\boldsymbol{\Xi}:{\rm Tr}(\boldsymbol{\Xi})=\mathcal{P}_L}\left\{BT\log\left[\det\left(\mathbf{I} + \mathbf{H}_{\Omega_L}\boldsymbol{\Xi}\mathbf{H}_{\Omega_L}^{\dagger}\right)\right]\right\} \tag{12}$$

for $L = 1,2,\ldots,K_{\rm bs}$ under given $\mathbf{H}$, where $(\cdot)^{\dagger}$ represents the conjugate transpose, $\det(\cdot)$ generates the determinant of a matrix, ${\rm Tr}(\cdot)$ evaluates the trace of a matrix, and $\boldsymbol{\Xi}$ is the covariance matrix of $\mathbf{s}_{\Omega_L}$. Also, we define $R(\Omega_L) \triangleq 0$ for $L = 0$, implying that no BS is selected and thus no data is transmitted in this fading state. For the details on how to achieve MIMO capacity under the given power budget and how to allocate power across multiple transmit antennas, please refer to [5].

After obtaining $R(\Omega_L)$, we have the best selection strategy to optimize the achievable rate as follows:

$$\max_{\Omega_L}\left\{R(\Omega_L)\right\}. \tag{13}$$

00. Let $\Psi := \{1, 2, \ldots, K_{\rm bs}\}$ and $\overline{\Psi} := \varnothing$, and $Z = |\Psi|$, where $\varnothing$ is the empty set and $|\Psi|$ denotes the cardinality of the set $\Psi$;<br>
! Use variables $\overline{\Psi}$ and $\Psi$ to store all selected BS's and all other BS's, respectively.<br>
01. For $i := 1$ to $L$ ! Add one BS to $\overline{\Psi}$ in each step.<br>
02. For $z := 1$ to $Z$ ! Examine $Z$ BS's in $\Psi$, respectively<br>
03. $\Theta_z := \overline{\Psi} \cup \{\psi_z\}$, where $\psi_z$ is the $z$th element in $\Psi$;<br>
! Pick a BS of $\Psi$ to form a new BS subset $\Theta_z$ with $\overline{\Psi}$.<br>
04. $\widetilde{R}_z := R(\Theta_z)$ based on Eq. (12) by setting $\Omega_L := \Theta_z$.<br>
! Examine the achievable rate of $\Theta_z$.<br>
05. End<br>
06. $z^* := \arg\max_{1 \le z \le Z} \left\{\widetilde{R}_z\right\}$ ;<br>
! Select the BS to maximize the achievable rate.<br>
07. $\overline{\Psi} := \Theta_{z^*}$, $\Psi := \Psi \backslash \{\psi_{z^*}\}$, and $Z := |\Psi|$;<br>
! Add the newly selected BS into the BS subset $\overline{\Psi}$.<br>
08. End<br>
09. $\Omega_L := \overline{\Psi}$. ! Complete the BS selection and get $\Omega_L$.

Fig. 3. The pseudo codes to determine $\Omega_L$ by using the incremental BS-selection algorithm for the single-user case.

To derive the optimal solution for this optimization problem, it is clear that we need to examine all $\binom{K_{\rm bs}}{L}$ possible BS combinations, which leads to the prohibitively high computational complexity as $M$ gets large. Alternatively, we consider two suboptimal approaches with low complexities as follows.

*1). Incremental BS-Selection Algorithm:* In [7], the authors developed the fast antenna selection algorithm using the incremental-selection strategy. Although this incremental-selection strategy was developed for antenna selection without CSI feedback, it can be readily extended to the scenario for BS selection with CSI feedback to achieve the near optimal data rate. The pseudo codes of the incremental BS-selection algorithm are given in Fig. 3. In particular, the idea of this algorithm is to determine $\Omega_L$ through $L$ steps, where in each step one BS is selected, as shown in lines 01-08 of Fig. 3. In each step, one selected BS is added to the BS subset denoted by $\overline{\Psi}$, where the selection criterion is to maximize the achievable rate of the updated BS subset $\overline{\Psi}$. Then, after $L$ steps, we have totally added $L$ BS's into $\overline{\Psi}$ and then assign $\Omega_L := \overline{\Psi}$. This algorithm only examines the achievable rates for $L(K_{\rm bs} - (L-1)/2)$ different BS combinations, which requires $L(K_{\rm bs} - (L-1)/2)$ times of singular value decomposition (SVD) to calculate the MIMO channel capacity, resulting in much less complexity than the optimal approach which examines all $\binom{K_{\rm bs}}{L}$ BS combinations.

*2). Ordered-Gain Based BS-Selection Algorithm:* The ordered-gain (or ordered-SNR) based BS-selection algorithm selects $L$ BS's with the largest aggregate power gain over all BS's, where the aggregate power gain is defined by Eq. (1). Since maximizing the aggregate power gain may not effectively optimize the achievable transmission rate for MIMO links, the incremental BS-selection algorithm usually dominates the ordered-gain based BS-selection algorithm. However, since the ordered-gain based BS-selection algorithm does not need to perform the SVD, its complexity is much lower than that of the incremental BS-selection algorithm.

### B. Time Sharing and Probabilistic Transmissions

To get the more general framework for BS selection, we apply the time sharing and probabilistic transmission strategies, respectively, over different transmission modes, which are described as follows.

*1). Time Sharing Transmissions:* Each time frame can be divided into $(K_{\rm bs}+1)$ time slots with the lengths equal to $\{T\alpha_m\}_{m=0}^{K_{\rm bs}}$, where $\alpha_m$ is the normalized time-slot length and $\sum_{m=0}^{K_{\rm bs}} \alpha_m = 1$. Within the $m$th time slot for $m > 1$, the transmission mode $L$ BS will be used; for $m = 0$, no data is transmitted in the corresponding time slot. Then, the total service rate in a time frame is equal to $\sum_{L=0}^{K_{\rm bs}} \alpha_L R(\Omega_L)$, where $R(\Omega_L)$ is given by Eq. (12) for $L \neq 0$ and $R(\Omega_L) = 0$ for $L = 0$. Furthermore, the total BS usage is given by $\sum_{L=0}^{K_{\rm bs}} L\alpha_L$. The purpose of applying the time sharing based transmissions is to increase the system flexibility and to gain the continuous control on the BS usage within each time frame. Accordingly, we need to identify how to optimally adjust $\boldsymbol{\alpha}$ with CSI and QoS constraints, where $\boldsymbol{\alpha} \triangleq (\alpha_0, \alpha_1, \ldots, \alpha_M)$.

*2). Probabilistic Transmissions:* Under this strategy, within each time frame only one transmission mode will be used for distributed MIMO transmissions. In particular, we will select transmission mode $L$ with probability equal to $\phi_L$ and define $\boldsymbol{\phi} \triangleq (\phi_0, \phi_1, \phi_2, \ldots, \phi_{K_{\rm bs}})^\tau$. Then, our target is to determine how to dynamically adjust $\boldsymbol{\phi}$ according to the CSI and QoS requirements.

If setting $\boldsymbol{\phi} = \boldsymbol{\alpha}$ and using the same strategy to determine $\Omega_L$ over all fading states, we obtain the same BS usage. However, the effective capacities (see Eq. (10)) achieved under the time-sharing transmission and the probabilistic transmission, denoted by $\mathcal{C}_{\rm TS}(\boldsymbol{\alpha}, \theta_1)$ and $\mathcal{C}_{\rm PR}(\boldsymbol{\phi}, \theta_1)$, respectively, are different. Specifically, we derive

$$\begin{aligned}
\mathcal{C}_{\rm TS}(\boldsymbol{\alpha}, \theta_1) &= -\frac{1}{\theta_1} \log\left(\mathbb{E}_{\mathbf{H}}\left\{e^{-\theta_1 \sum_{L=0}^{K_{\rm bs}} \alpha_L R(\Omega_L)}\right\}\right) \\
&\geq -\frac{1}{\theta_1} \log\left(\mathbb{E}_{\mathbf{H}}\left\{\alpha_L e^{-\theta_1 \sum_{L=0}^{K_{\rm bs}} R(\Omega_L)}\right\}\right) \\
&= -\frac{1}{\theta_1} \log\left(\mathbb{E}_{\mathbf{H}}\left\{\phi_L e^{-\theta_1 \sum_{L=0}^{K_{\rm bs}} R(\Omega_L)}\right\}\right) \\
&= \mathcal{C}_{\rm PR}(\boldsymbol{\phi}, \theta_1), \qquad (14)
\end{aligned}$$

where the inequality holds because $\mathbb{E}_{\mathbf{H}}\left\{e^{-\sum_{L=0}^{K_{\rm bs}} \alpha_L R(\Omega_L)}\right\}$ is convex over $(R(\Omega_0), R(\Omega_1), \ldots, R(\Omega_{K_{\rm bs}}))$. Eq. (14) suggests that the time-sharing transmission generally outperforms the probabilistic transmission. However, the probabilistic transmission is more realistic to implement than the time-sharing transmission due to the following reasons. On the one hand, for the optimized time-sharing transmission, the time-slot length $T\alpha_L$ may be quite small and thus very hard to implement. On the other hand, the multiple time slots (for the time-sharing transmission) within a time frame introduces more overhead than the single-slot case (for the probabilistic transmission).

### C. Optimization Framework Using Time-Sharing Transmissions with Incremental BS Selection

As discussed in Section II-C, our major objective is to minimize the average BS usage. In this section, we focus on the framework which employs the incremental BS-selection algorithm for each transmission mode and apply the time-sharing transmission for different transmission modes. Then, we develop the efficient BS-selection scheme under the above framework by solving the following optimization problem $\boldsymbol{A1}$, which aims at minimizing the average BS usage while guaranteeing the delay-QoS requirement.

$$\boldsymbol{A1}: \min_{\boldsymbol{\alpha}(\mathbf{H})} \left\{\overline{L}\right\} = \min_{\boldsymbol{\alpha}(\mathbf{H})} \left\{\mathbb{E}_{\mathbf{H}}\left\{\sum_{L=0}^{K_{\rm bs}} \alpha_L(\mathbf{H})L\right\}\right\}$$

$$\text{s.t.: 1). } \sum_{L=0}^{K_{\rm bs}} \alpha_L(\mathbf{H}) = 1, \quad \forall \mathbf{H}; \tag{15}$$

$$\text{2). } \mathbb{E}_{\mathbf{H}}\left\{e^{-\theta_1 \sum_{L=0}^{K_{\rm bs}} \alpha_L(\mathbf{H})R(\Omega_L)} - e^{-\theta_1 \overline{C}_1}\right\} \leq 0, \tag{16}$$

where Eq. (16) is the constraint to guarantee the delay-bound violation probability as derived in Eq. (2), we obtain $\Omega_L$ through the incremental selection algorithm listed in Fig. 2, which varies with $\mathbf{H}$, and we determine $R(\Omega_L)$ based on Eq. (12). Note that we denote the time-sharing vector (originally defined in Section IV-B) by $\boldsymbol{\alpha}(\mathbf{H})$ in problem $\boldsymbol{A1}$ to emphasize that $\boldsymbol{\alpha}$ is a function of $\mathbf{H}$. When the context is clear, we will then drop the symbol $(\mathbf{H})$ and only use $\boldsymbol{\alpha}$ to simplify notations without causing confusions. We call the optimal solution to $\boldsymbol{A1}$, denoted by $\boldsymbol{\alpha}^*$, as the *incremental BS-selection and time-sharing based* (IBS-TS) scheme.

Note that the optimization over $\boldsymbol{\alpha}$ is an $(K_{\rm bs}+1)$-dimensional problem. To reduce the dimension of optimization variables, for each $\mathbf{H}$ we define

$$\widetilde{R}(\mathcal{L}) \triangleq \max_{\boldsymbol{\alpha}} \left\{\sum_{L=0}^{K_{\rm bs}} \alpha_L R(\Omega_L)\right\}, \qquad \text{for each } \mathbf{H} \tag{17}$$

$$\text{s.t. } \sum_{L=0}^{K_{\rm bs}} \alpha_L = 1; \ \sum_{L=0}^{K_{\rm bs}} \alpha_L L = \mathcal{L}, \ \text{for each } \mathbf{H}. \tag{18}$$

Based on Eqs. (17)-(18), $\widetilde{R}(\mathcal{L})$ is the maximum achievable rate over all $\boldsymbol{\alpha}$ with the same BS usage. Since $\widetilde{R}(\mathcal{L})$ is the convex combination [12] over $\{R(\Omega_L)\}_{L=1}^{K_{\rm bs}}$, then the definition in Eqs. (17)-(18) suggests that $\widetilde{R}(\mathcal{L})$ is a piece-wise linear and concave function, which can be written as

$$\widetilde{R}(\mathcal{L}) = \begin{cases} \widetilde{R}(m_{j-1}) + \nu_j(\mathcal{L} - m_{j-1}), & \text{if } \mathcal{L} \in (m_{j-1}, m_j], \\ & j = 1, 2, \dots, \mathcal{K}; \\ 0, & \text{if } \mathcal{L} = 0; \end{cases} \tag{19}$$

for a certain integer $\mathcal{K}$, where $m_0 < m_1 < \cdots < m_{\mathcal{K}}$, $m_0 = 0$, $m_{\mathcal{K}} = K_{\rm bs}$, and $m_i \in \{0, 1, \dots, K_{\rm bs}\}$. Moreover, using $\big(m_i, \widetilde{R}(m_i)\big)$ to represent the coordinates of a point in the two-dimensional plane, we can identify $\big\{\big(m_i, \widetilde{R}(m_i)\big)\big\}_{i=1}^{\mathcal{K}-1}$ through the following procedures: a). find vertices of the convex hull spanned by two-dimensional points $\big\{\big(L, \widetilde{R}(L)\big)\big\}_{i=0}^{K_{\rm bs}}$; b). $\big\{\big(m_i, \widetilde{R}(m_i)\big)\big\}_{i=1}^{\mathcal{K}-1}$ are located above the line segment with end points $(0,0)$ and $\big(K_{\rm bs}, \widetilde{R}(K_{\rm bs})\big)$. Accordingly, $\nu_j$ is the slope of the line segment starting at the point $\big(m_i, \widetilde{R}(m_i)\big)$ and ending at the point $\big(m_{i-1}, \widetilde{R}(m_{i-1})\big)$, which is determined by

$$\nu_i = \frac{\widetilde{R}(m_i) - \widetilde{R}(m_{i-1})}{m_i - m_{i-1}}, \qquad i = 1, 2, \dots, \mathcal{K}. \tag{20}$$

For presentation convenience, we also define $\nu_0 \triangleq \infty$ and $\nu_{\mathcal{K}+1} \triangleq -\infty$. Note that $\widetilde{R}(\mathcal{L})$ defined in Eq. (19) and the associated variables, including $\Omega_L$, $\mathcal{K}$, $\{m_j\}_{j=1}^{\mathcal{K}}$, and $\{\nu_i\}_{i=1}^{\mathcal{K}}$, are all functions of $\mathbf{H}$. Furthermore, given $\mathcal{L} \in [m_{j-1}, m_j]$, following the piece-wise linear property, we derive the corresponding $\boldsymbol{\alpha}$ to achieve the service rate $\widetilde{R}(\mathcal{L})$ as follows:

$$\alpha_L = \begin{cases} \frac{m_j - \mathcal{L}}{m_j - m_{j-1}}, & \text{if } L = m_{j-1}; \\ \frac{\mathcal{L} - m_{j-1}}{m_j - m_{j-1}}, & \text{if } L = m_j; \\ 0, & \text{otherwise.} \end{cases} \tag{21}$$

Applying Eqs. (17), (18), and (21) into problem $\boldsymbol{A1}$, we can equivalently convert $\boldsymbol{A1}$ to the following problem $\boldsymbol{A1}'$:

$$\boldsymbol{A1}': \ \min_{\mathcal{L}(\mathbf{H})} \left\{\mathbb{E}_{\mathbf{H}}\left\{\mathcal{L}(\mathbf{H})\right\}\right\}$$

$$\text{s.t.: } \ \mathbb{E}_{\mathbf{H}}\left\{e^{-\theta_1 \widetilde{R}(\mathcal{L}(\mathbf{H}))} - e^{-\theta_1 \overline{C}_1}\right\} \leq 0, \tag{22}$$

where $\mathcal{L}(\mathbf{H})$ is a function of $\mathbf{H}$ and we denote the optimal solution to problem $\boldsymbol{A1}'$ by $\mathcal{L}^*(\mathbf{H})$. To simplify notations, we will drop the symbol $\mathbf{H}$ for problem $\boldsymbol{A1}'$. After obtaining the optimal solution $\mathcal{L}^*$, we can then uniquely map $\mathcal{L}^*$ to $\boldsymbol{\alpha}$ through Eq. (21). Since $\widetilde{R}(\mathcal{L})$ is an increasing and concave function, $e^{-\theta_1 \widetilde{R}(\mathcal{L})}$ is convex over $\mathcal{L}$ [14, pp. 84]. Thus, we can see that $\boldsymbol{A1}'$ satisfies: a) the objective function is convex; b) the inequality constraint function $\mathbb{E}_{\mathbf{H}}\big\{e^{-\theta_1 \widetilde{R}(\mathcal{L})} - e^{-\theta_1 \overline{C}_1}\big\}$ is convex. Therefore, $\boldsymbol{A1}'$ is a convex problem [14, pp. 137]. Then, using the Lagrangian method, we solve for the optimal solution of $\boldsymbol{A1}'$, as summarized in the following Theorem 1.

*Theorem 1:* The optimal solution to $\boldsymbol{A1}'$, if existing, is determined by

$$\mathcal{L}^* = \begin{cases} m_j, & \text{if } \nu_{j+1} \leq \dfrac{e^{\theta_1 \widetilde{R}(m_j)}}{\theta_1 \lambda^*} \leq \nu_j; \\ \dfrac{\log(\theta_1 \lambda^* \nu_j)}{\nu_j \theta_1} - \dfrac{\widetilde{R}(m_{j-1})}{\nu_j} + m_{j-1}, & \\ & \text{if } \widetilde{R}(m_{j-1}) < \frac{\log(\theta_1 \lambda^* \nu_j)}{\theta_1} < \widetilde{R}(m_j), \end{cases} \tag{23}$$

where $\widetilde{R}(\cdot)$ and $\nu_j$ are characterized by Eqs. (17) through (20). In Eq. (23), $\lambda^* \geq 0$ is a constant over all fading states, which needs to be selected such that equality in Eq. (16) holds.

*Proof:* The proof of Theorem 1 is provided in the Appendix A. ■

*Remarks:* (i) Having obtained the optimal $\mathcal{L}^*$ for problem $\boldsymbol{A1}'$, the optimal solution to $\boldsymbol{A1}$, denoted by $\boldsymbol{\alpha}^*$, is obtained by setting $\mathcal{L} = \mathcal{L}^*$ in Eq. (21). (ii) Under the optimal solution, we do not allocate time slots for all transmission modes. As indicated by Eq. (21), within any time frames, we employ at most two transmission modes. (iii) It is clear that by setting $\alpha_{K_{\rm bs}} = 1$ for all time frames, we use the maximum transmit power and thus obtain the maximum achievable effective capacity, which is denoted by $\mathcal{C}_{\max}^{(1)}$. If $\mathcal{C}_{\max}^{(1)}$ is still smaller than $\overline{C}_1$, the specified delay-QoS requirement cannot be satisfied since we have used up all power budget. As a result, no feasible solution exists for this case. In contrast, if $\mathcal{C}_{\max}^{(1)} \geq \overline{C}_1$, we can always find the optimal solution. (iv) The analytical expressions for $\lambda^*$ is usually intractable. Alternatively, we use the following numerical-tracking method to derive the optimal $\lambda^*$. The details of the numerical-tracking method and the discussions on its complexity are provided in Appendix B.

### *D. Optimization Framework Using Probabilistic Transmissions with Ordered-Gain Based BS Selection*

We in this section consider the framework using the ordered-gain based BS-selection algorithm and the probabilistic transmission strategy. Specifically, we formulate the optimization problem for this framework as follows:

$$\boldsymbol{A2}: \min_{\boldsymbol{\phi}(\mathbf{H})} \{\overline{L}\} = \min_{\boldsymbol{\phi}(\mathbf{H})} \left\{ \mathbb{E}_{\mathbf{H}} \left\{ \sum_{L=0}^{K_{\rm bs}} \phi_L(\mathbf{H}) L \right\} \right\}$$

$$\text{s.t.: } 1).\ \sum_{L=0}^{K_{\rm bs}} \phi_L(\mathbf{H}) = 1, \quad \forall \mathbf{H}; \tag{24}$$

$$2).\ \mathbb{E}_{\mathbf{H}} \left\{ \left( \sum_{L=0}^{K_{\rm bs}} \phi_L(\mathbf{H}) e^{-\theta_1 R(\Omega_L)} \right) - e^{-\theta_1 \overline{C}_1} \right\} \le 0, \tag{25}$$

where the probabilistic transmission vector is denoted by a function $\boldsymbol{\phi}(\mathbf{H})$ of $\mathbf{H}$. When the context is clear, we drop the symbol $(\mathbf{H})$ for problem $\boldsymbol{A2}$ to simplify notations. We call the optimal solution to the optimization problem $\boldsymbol{A2}$ as the ordered-gain and probability transmission based (OGBS-PT) scheme.

*Theorem 2:* The optimal solution to problem $\boldsymbol{A2}$, denoted by $\boldsymbol{\phi}^*$, is given by

$$\phi_L^* = \begin{cases} 1, & \text{if } L = L^*; \\ 0, & \text{if } L \neq L^*, \end{cases} \tag{26}$$

for all $\mathbf{H}$, where

$$L^* = \arg\min_L \left\{ L + \lambda^* e^{-\theta_1 R(\Omega_L)} \right\}. \tag{27}$$

In Eq. (27), $\lambda^* \ge 0$ is a constant over all $\mathbf{H}$, which is selected such that the equality holds for Eq. (25) to guarantee the delay-QoS requirement.

*Proof:* The detailed proof of Theorem 2 is omitted due to lack of space, but is provided on-line in the technical-report version [19] of this paper. ■

*Remarks:* Theorem 2 suggests that under the optimal solution, the probabilistic transmission reduces to a deterministic strategy, where the only transmission mode $L^*$ will be used for data transmission. Similar to problem $\boldsymbol{A1}$, the parameter $\lambda^*$ for $\boldsymbol{A2}$ also needs to be tracked through numerical searching.

## V. QoS-Aware BS Selection for the Multi-User Case

We next consider the distributed MIMO transmissions for the case with *multiple mobile users*.[1] For efficient BS selection and distributed MIMO transmissions, the central server controls the selected distributed BS's and the mobile users to constitute the broadcast MIMO link, as mentioned in Section II. Specifically, given the transmission mode[2] $L$ and BS-index subset $\Omega_L = \{i_{L,1}, i_{L,2}, \ldots, i_{L,L}\}$ of the selected BS's, the channel matrix of the $n$th mobile user, modeled by $\mathbf{H}_{\Omega_L}^{(n)}$, is determined by

$$\mathbf{H}_{\Omega_L}^{(n)} \triangleq \left[ \mathbf{H}_{n,i_{L,1}}\ \mathbf{H}_{n,i_{L,2}}\ \cdots\ \mathbf{H}_{n,i_{L,L}} \right],\ n = 1, 2, \ldots, K_{\rm mu}$$

where $\mathbf{H}_{\Omega_L}^{(n)}$ is an $N_n \times \left( \sum_{\ell=1}^{L} M_{i_{L,\ell}} \right)$ matrix. Then, the physical-layer signal transmissions can be characterized by

$$\mathbf{y}_{\Omega_L}^{(n)} = \mathbf{H}_{\Omega_L}^{(n)} \sum_{i=1}^{K_{\rm mu}} \mathbf{s}_{\Omega_L}^{(i)} + \boldsymbol{\varsigma}^{(n)}, \quad n = 1, 2, \ldots, K_{\rm mu}$$

where $\mathbf{s}_{\Omega_L}^{(i)}$ represents the $i$th user's input signal vector for the MIMO channel $\mathbf{H}_{\Omega_L}^{(n)}$, $\mathbf{y}_{\Omega_L}^{(n)}$ is the signal vector received at the $n$th user's receive antennas, and $\boldsymbol{\varsigma}^{(n)}$ is the complex additive white Gaussian noise (AWGN) vector with unit power for each element of this vector.

It is well-known that the optimal capacity of the broadcast MIMO link can be achieved through the dirty-paper coding techniques [16], which is, however, hard to implement due to its high complexity [17]. Alternatively, we apply the block-diagonalization precoding techniques [17] for distributed MIMO transmissions and then concentrate on developing efficient QoS-aware BS-selection schemes and associated resource allocation schemes, which are elaborated on in the following sections.

### *A. The Block Diagonalization Technique for Distributed MIMO Transmissions*

The idea of block diagonalization (BD) [17] is to use a precoding matrix, denoted by $\mathbf{\Gamma}_{\Omega_L}^{(n)}$, for the $n$th user's transmitted signal vector, such that $\mathbf{H}_{\Omega_L}^{(i)} \mathbf{\Gamma}_{\Omega_L}^{(n)} = \mathbf{0}$ for all $i \neq n$. By setting $\mathbf{s}_{\Omega_L}^{(n)} = \mathbf{\Gamma}_{\Omega_L}^{(n)} \widehat{\mathbf{s}}_{\Omega_L}^{(n)}$, where $\widehat{\mathbf{s}}_{\Omega_L}^{(n)}$ is the $n$th user's data vector to be precoded by $\mathbf{\Gamma}_{\Omega_L}^{(n)}$, we can rewrite the received signal $\mathbf{y}_{\Omega_L}^{(n)}$ as

$$\begin{aligned} \mathbf{y}^{(n)} &= \mathbf{H}_{\Omega_L}^{(n)} \sum_{i=1}^{K_{\rm mu}} \mathbf{\Gamma}_{\Omega_L}^{(i)} \widehat{\mathbf{s}}_{\Omega_L}^{(i)} + \boldsymbol{\varsigma}^{(n)} = \mathbf{H}_{\Omega_L}^{(n)} \mathbf{\Gamma}_{\Omega_L}^{(n)} \widehat{\mathbf{s}}_{\Omega_L}^{(n)} + \boldsymbol{\varsigma}^{(n)} \\ &= \widehat{\mathbf{\Gamma}}_{\Omega_L}^{(n)} \widehat{\mathbf{s}}_{\Omega_L}^{(n)} + \boldsymbol{\varsigma}^{(n)}, \end{aligned} \tag{28}$$

where $\widehat{\mathbf{\Gamma}}_{\Omega_L}^{(n)} \triangleq \mathbf{H}_{\Omega_L}^{(n)} \mathbf{\Gamma}_{\Omega_L}^{(n)}$. Under this strategy, the $n$th user's signal will not cause interferences to other users. Accordingly, the MIMO broadcast transmissions are virtually converted to $K_{\rm mu}$ orthogonal MIMO channels with channel matrices $\{\widehat{\mathbf{\Gamma}}_{\Omega_L}^{(n)}\}_{n=1}^{K_{\rm mu}}$. Then, we can get the $n$th user's maximum achievable rate in a fading state, denoted by $R^{(n)}\big(\Omega_L, \mathcal{P}_L^{(n)}\big)$, as follows:

$$\begin{aligned} & R^{(n)}\big(\Omega_L, \mathcal{P}_L^{(n)}\big) \\ & \triangleq \max_{\mathbf{\Xi}^{(n)}} \left\{ BT \log \left[ \det \left( \mathbf{I} + \widehat{\mathbf{\Gamma}}_{\Omega_L}^{(n)} \mathbf{\Xi}^{(n)} \left( \widehat{\mathbf{\Gamma}}_{\Omega_L}^{(n)} \right)^{\dagger} \right) \right] \right\}, \\ & \quad \text{s.t.: } \mathrm{Tr}\left( \mathbf{\Xi}^{(n)} \right) = \mathcal{P}_L^{(n)}, \end{aligned} \tag{29}$$

where $\mathbf{\Xi}^{(n)}$ is the covariance matrix of $\widehat{\mathbf{s}}_{\Omega_L}^{(n)}$ and $\mathcal{P}_L^{(n)}$ is the power allocated for the $n$th user. Also, we define

[1] We use the terms of "mobile user" and "user" exchangeably in the rest of this paper.

[2] For the multi-user case, we also use the term of transmission mode $L$ to denote the case where the BS-subset's cardinality is $L$.

$R^{(n)}\big(\Omega_L, \mathcal{P}_L^{(n)}\big) \triangleq 0$ for $L=0$. Note that the aforementioned $\mathbf{\Gamma}_{\Omega_L}^{(n)}$ may not exist, implying that the total number of transmit antennas over the selected BS's cannot support enough freedom for block-diagonalization. In this case, no data will be transmitted to the $n$th user to avoid interferences caused to other users. For the procedures on how to determine $\mathbf{\Gamma}_{\Omega_L}^{(n)}$, please refer to [17].

### *B. Priority BS-Selection Strategy Given BS Subset Cardinality*

When the transmission mode is specified, i.e., the cardinality of the BS subset is given, every user expects to select the BS subset that maximizes its own transmission rate. However, it is clear that this objective cannot be obtained for all users in the multi-user case. Moreover, the derivation of global optimal selection strategy in terms of minimizing the average BS usage is intractable due to the too high complexity, where we need to examine all $\binom{K_{\rm bs}}{L}$ possible BS combinations. Therefore, we propose a simple yet efficient BS-selection algorithm, called priority BS-selection (PBS), which is detailed as follows.

For the $n$th user, the global maximum achievable transmission rate is attained when all BS's are used and all the other users do not transmit. Thus, the maximum achievable rate is given by

$$R^{(n)}\big(\Omega_{K_{\rm bs}}, \mathcal{P}_{K_{\rm bs}}\big) = \max_{\mathbf{\Xi}^{(n)}:{\rm Tr}\left(\mathbf{\Xi}^{(n)}\right)=\mathcal{P}_{K_{\rm bs}}} \left\{BT\log\Big[\det\Big(\mathbf{I}+\mathbf{H}_n\mathbf{\Xi}^{(n)}\mathbf{H}_n^{\dagger}\Big)\Big]\right\}. \quad (30)$$

Then, we get the maximum achievable effective capacity of the $n$th user, denoted by $\mathcal{C}_{\max}^{(n)}$, as follows:

$$\mathcal{C}_{\max}^{(n)} = -\frac{1}{\theta_n}\log\left(\mathbb{E}_{\mathbf{H}}\left\{e^{-\theta_n R^{(n)}\left(\Omega_{K_{\rm bs}},\mathcal{P}_{K_{\rm bs}}\right)}\right\}\right) \quad (31)$$

for $n=1,2,\dots,K_{\rm mu}$. We further define the effective-capacity fraction for the $n$th user as the ratio between the traffic loads and the maximum achievable effective capacity. Denoting the effective-capacity fraction by $\widehat{C}_n$, we have $\widehat{C}_n \triangleq \overline{C}_n/\mathcal{C}_{\max}^{(n)}$. Clearly, for a higher $\widehat{C}_n$, the $n$th user needs more wireless resources to meet its QoS requirements. Thus, in order to satisfy the QoS requirements for all users, we assign higher BS-selection priority to the user with larger $\widehat{C}_n$. Following this principle, we design the priority BS-selection algorithm to determine $\Omega_L$ in each fading state and provide the pseudo code in Fig. 4. For presentation convenience, we sort $\{\widehat{C}_n\}_{n=1}^{K_{\rm mu}}$ in the decreasing order and denote the permuted version by $\{\widehat{C}_{\pi(j)}\}_{h=1}^{K_{\rm mu}}$, where $\widehat{C}_{\pi(1)} \ge \widehat{C}_{\pi(2)} \ge \cdots \ge \widehat{C}_{\pi(K_{\rm mu})}$ indicates the order from the higher priority to the lower priority. In the rest of this paper, we use the term of user $\pi(i)$ to denote the user with the $i$th largest effective-capacity fraction.

As shown in Fig. 4, in each fading state the BS-selection procedure starts with the selection for user $\pi(1)$, who has the highest priority. After picking one BS for user $\pi(1)$, we select one different BS for user $\pi(2)$. More generally, after selecting for user $\pi(j)$, we choose one BS for user $\pi(j+1)$ from the BS-subset $\Psi$, which consists of the BS's that have not been selected. This procedure repeats until $L$ BS's are selected. For user-$\pi(j)$'s selection, we choose the BS with the maximum aggregate power gain (see Eq. (1) for its definition) over the

01. Let $\Psi := \{1,2,\dots,K_{\rm bs}\}$ and $\overline{\Psi} := \varnothing$, and $\ell = |\overline{\Psi}|$;
    ! Use variables $\overline{\Psi}$ and $\Psi$ to store selected BS's and all other BS's, respectively.
02. $j := 1$. ! User $\pi(j)$ is selecting BS.
03. While $(\ell < L)$ ! Iterative selections until $L$ BS's are selected.
04. $m^* = \arg\min_{m\in\Psi}\{\gamma_{\pi(j),m}\}$.
    ! $\gamma_{\pi(j),m}$ is the aggregate power gain associated with user $\pi(j)$.
    ! Select BS to maximum the aggregate power gain for user $\pi(j)$.
05. $\overline{\Psi} := \overline{\Psi}\cup\{m^*\}$, $\Psi := \Psi\backslash\{m^*\}$, and $\ell := \ell+1$.
    ! Update $\overline{\Psi}$, $\Psi$, and $\ell$.
06. If $j = K_{\rm mu}$, then $j := 1$; else $j := j+1$.
    ! Let next user with lower priority to select BS.
07. End
08. $\Omega_L := \overline{\Psi}$. ! Complete the BS selection and get $\Omega_L$.

Fig. 4. The pseudo codes to determine $\Omega_L$ in each fading state by using the priority BS-selection algorithm for the multi-user case.

subset $\Psi$. In addition, after user-$\pi(K_{\rm mu})$'s selection, if the number of selected BS's is still smaller than $L$, we continue selecting one more BS for user $\pi(1)$, as shown in line 06 in Fig. 4, and repeat this iterative selection procedure until having selected $L$ BS's.

### *C. The Optimization Framework for BS-Selection and Resource Allocation*

#### *C.1. Problem Formulation for Average BS-Usage Minimization*

We next study how to determine which transmission modes will be used, and how to derive the corresponding resource allocation strategy by integrating the block diagonalization and the priority BS selection. Similar to the OGBS-PT scheme for the single-user case, we also apply the probabilistic transmission (PT) for the multi-user case, where transmission mode $L$ is used with a probability denoted by $\phi_L$, $L=1,2,\dots,K_{\rm bs}$. Note that for each transmission mode, there are $K_{\rm mu}$ coexisting links towards $K_{\rm mu}$ mobile users. Consequently, we also need to determine how to allocate the total power $\mathcal{P}_L$ to these $K_{\rm mu}$ coexisting links. In particular, we describe the power allocation strategy in a fading state by $\boldsymbol{\mathcal{P}} \triangleq \big(\boldsymbol{\mathcal{P}}_1, \boldsymbol{\mathcal{P}}_2, \dots, \boldsymbol{\mathcal{P}}_{K_{\rm bs}}\big)$ with $\boldsymbol{\mathcal{P}}_L \triangleq \big(\mathcal{P}_L^{(1)}, \mathcal{P}_L^{(2)}, \dots, \mathcal{P}_L^{(K_{\rm bs})}\big)$ for $L=1,2,\dots,K_{\rm bs}$, where $\boldsymbol{\mathcal{P}}$ denotes the power-allocation policy for the entire system and the vector $\boldsymbol{\mathcal{P}}_L$ represents the power-allocation policy for transmission mode $L$. Then, we formulate the following optimization problem $\boldsymbol{A3}$ to derive the optimal QoS-aware probability-vector $\boldsymbol{\phi}^* \triangleq (\phi_1^*, \phi_2^*, \dots, \phi_{K_{\rm mu}}^*)$ and its corresponding power-allocation policy $\boldsymbol{\mathcal{P}}^*$.

$$\boldsymbol{A3}: \min_{(\boldsymbol{\phi}(\mathbf{H}),\boldsymbol{\mathcal{P}}(\mathbf{H}))}\{\overline{L}\} = \min_{(\boldsymbol{\phi}(\mathbf{H}),\boldsymbol{\mathcal{P}}(\mathbf{H}))}\left\{\mathbb{E}_{\mathbf{H}}\left\{\sum_{L=0}^{K_{\rm bs}}\phi_L(\mathbf{H})L\right\}\right\}$$

$$\text{s.t.: } 1).\ \sum_{L=0}^{K_{\rm bs}}\phi_L(\mathbf{H}) = 1, \qquad \forall\,\mathbf{H}; \quad (32)$$

$$2).\ \sum_{n=1}^{K_{\rm mu}}\mathcal{P}_L^{(n)}(\mathbf{H}) = \mathcal{P}_L, \qquad \forall\,L,\,\mathbf{H}; \quad (33)$$

$$3).\ \mathbb{E}_{\mathbf{H}}\left\{\sum_{L=0}^{K_{\rm bs}}\phi_L(\mathbf{H})e^{-\theta_n R^{(n)}\left(\Omega_L,\mathcal{P}_L^{(n)}(\mathbf{H})\right)}\right\} \le e^{-\theta_n\overline{C}_n}, \quad \forall\,n, \quad (34)$$

where $\boldsymbol{\phi}(\mathbf{H})$ and $\boldsymbol{\mathcal{P}}(\mathbf{H})$ characterize the probabilistic transmission vector and the power allocation policy described in the beginning of Section V.C.1, both of which vary with $\mathbf{H}$, and $R^{(n)}\big(\Omega_L, \mathcal{P}_L^{(n)}(\mathbf{H})\big)$ is determined through Eq. (29). After clearly define $\boldsymbol{\phi}(\mathbf{H})$ and $\boldsymbol{\mathcal{P}}(\mathbf{H})$, we drop the symbol $(\mathbf{H})$ in problem $\boldsymbol{A3}$ to simplify notations. We call the optimal solution to $\boldsymbol{A3}$ as the PBS-BD-PT scheme.

*C.2. The Properties of* $R^{(n)}\big(\Omega_L, \mathcal{P}_L^{(n)}\big)$

Before solving $\boldsymbol{A3}$, we need to study the properties of $R^{(n)}\big(\Omega_L, \mathcal{P}_L^{(n)}\big)$. Let us consider the $n$th user with $\mathbf{\Gamma}_{\Omega_L}^{(n)}$ not equal to zero. Similar to the results summarized in Section IV-A, the $n$th user's MIMO channel $\widehat{\mathbf{\Gamma}}_{\Omega_L}^{(n)}$ (after the block diagonalization) can be converted to $Z_L^{(n)}$ parallel Gaussian sub-channels, where $Z_L^{(n)}$ is the rank of $\widehat{\mathbf{\Gamma}}_{\Omega_L}^{(n)}$, the $z$th sub-channel's SNR is equal to $\varepsilon_{L,z}^{(n)}$, and $\sqrt{\varepsilon_{L,z}^{(n)}}$ is the $z$th largest nonzero singular value of $\widehat{\mathbf{\Gamma}}_{\Omega_L}^{(n)}$. The optimal power $\rho_{L,z}^{(n)}$ allocated to the $z$th sub-channel follows the water-filling allocation, which is equal to $\rho_{L,z}^{(n)} = \big[\mu_L^{(n)} - 1/\varepsilon_{L,z}^{(n)}\big]^+$, where $\mu_L^{(n)} \geq 1/\varepsilon_{L,1}^{(n)}$ and $\mu_L^{(n)}$ is selected such that $\sum_{z=1}^{Z_L^{(n)}} \rho_{L,z}^{(n)} = \mathcal{P}_L^{(n)}$. Since $\widehat{\mathbf{\Gamma}}_{\Omega_L}^{(n)}$ has only $Z_L^{(n)}$ non-zero singular values, for presentation convenience, we define $1/\varepsilon_{L,i}^{(n)} \triangleq \infty$ for $i = Z_L^{(n)} + 1$. Accordingly, we can show that

$$\frac{dR^{(n)}\big(\Omega_L, \mathcal{P}_L^{(n)}\big)}{d\mathcal{P}_L^{(n)}} = \frac{BT}{\mu_L^{(n)}} \tag{35}$$

holds and that $R^{(n)}\big(\Omega_L, \mathcal{P}_L^{(n)}\big)$ is strictly concave over $\mathcal{P}_L^{(n)}$. Moreover, if $\mu_L^{(n)} \in \Big[1/\varepsilon_{L,i}^{(n)}, 1/\varepsilon_{L,i+1}^{(n)}\Big)$ for $i = 1, 2, \ldots, Z_L^{(n)}$, we get:

$$\begin{cases} \text{(a).}\ \mathcal{P}_L^{(n)} = i\mu_L^{(n)} - \displaystyle\sum_{j=1}^{i} \frac{1}{\varepsilon_{L,j}^{(n)}} \\ \text{(b).}\ R^{(n)}\big(\Omega_L, \mathcal{P}_L^{(n)}\big) = BT \log\left(\displaystyle\prod_{j=1}^{i} \varepsilon_{L,j}^{(n)}\right) + BTi \log \mu_L^{(n)}. \end{cases} \tag{36}$$

*C.3. The Optimal Solution to* $\boldsymbol{A3}$

*Theorem 3:* The optimal power-allocation policy $\boldsymbol{\mathcal{P}}^*$ for optimization problem $\boldsymbol{A3}$, if existing, is given as follows:

$$\Big(\mathcal{P}_L^{(n)}\Big)^* = i^* \left(\prod_{j=1}^{i^*} \varepsilon_{L,j}^{(n)}\right)^{-\frac{BT\theta_n}{1+i^*BT\theta_n}} \left(\frac{\zeta_{\mathbf{H},L}^*}{BT\theta_n \lambda_n^*}\right)^{-\frac{1}{1+i^*BT\theta_n}} - \sum_{j=1}^{i^*} \frac{1}{\varepsilon_{L,j}^{(n)}}, \tag{37}$$

for all $n$, $L$, and $\mathbf{H}$, where $\varepsilon_{L,j}^{(n)}$ is the square of $\widehat{\mathbf{\Gamma}}_{\Omega_L}^{(n)}$'s $j$th largest singular value, and $i^*$ is the unique solution satisfying the following condition:

$$\mu_L^{(n)} \in \left[\frac{1}{\varepsilon_{L,i^*}^{(n)}}, \frac{1}{\varepsilon_{L,i^*+1}^{(n)}}\right), \quad \forall n,\ L,\ \mathbf{H}, \tag{38}$$

where

$$\mu_L^{(n)} = \max\left\{\frac{1}{\varepsilon_{L,1}^{(n)}}, \left(\prod_{j=1}^{i^*} \varepsilon_{L,j}^{(n)}\right)^{-\frac{BT\theta_n}{1+i^*BT\theta_n}} \left(\frac{\zeta_{\mathbf{H},L}^*}{BT\theta_n \lambda_n^*}\right)^{-\frac{1}{1+i^*BT\theta_n}}\right\}. \tag{39}$$

The corresponding optimal probability-transmission policy is determined by

$$\phi_L^* = \begin{cases} 1, & \text{if } L = L^*; \\ 0, & \text{otherwise} \end{cases} \tag{40}$$

with

$$L^* = \arg\min_{0 \leq L \leq K_{\text{bs}}} \left\{ L + \sum_{n=1}^{K_{\text{mu}}} \lambda_n^* e^{-\theta_n BT\left(\log\left(\prod_{j=1}^{i^*} \varepsilon_{L,j}^{(n)}\right) + i^* \log \mu_L^{(n)}\right)} \right\}, \ \forall \mathbf{H}, \tag{41}$$

where given $\{\lambda_n^*\}_{n=1}^{K_{\text{mu}}}$ for Eqs. (37)-(39), the optimal $\zeta_{\mathbf{H},L}^*$ is selected to satisfy the equation $\sum_{n=1}^{K_{\text{mu}}} \big(\mathcal{P}_L^{(n)}\big)^* = \mathcal{P}_L$ for all $L$ and $\mathbf{H}$; $\{\lambda_n^*\}_{n=1}^{K_{\text{mu}}}$ are constants, which are selected such that the equality of Eq. (34) holds.

*Proof:* We construct $\boldsymbol{A3}$'s Lagrangian function, denoted by $\mathcal{J}_{A3}(\boldsymbol{\phi}, \boldsymbol{\mathcal{P}}; \boldsymbol{\lambda}, \boldsymbol{\zeta}_{\mathbf{H}})$, as $\mathcal{J}_{A3}(\boldsymbol{\phi}, \boldsymbol{\mathcal{P}}; \boldsymbol{\lambda}, \boldsymbol{\zeta}_{\mathbf{H}}) = \mathbb{E}_{\mathbf{H}}\left\{J_{A3}(\boldsymbol{\phi}, \boldsymbol{\mathcal{P}}; \boldsymbol{\lambda}, \boldsymbol{\zeta}_{\mathbf{H}})\right\}$ with

$$J_{A3}(\boldsymbol{\phi}, \boldsymbol{\mathcal{P}}; \boldsymbol{\lambda}, \boldsymbol{\zeta}_{\mathbf{H}}) \triangleq \sum_{n=1}^{K_{\text{mu}}} \lambda_n \left[\sum_{L=0}^{K_{\text{bs}}} \phi_L e^{-\theta_n R^{(n)}\left(\Omega_L, \mathcal{P}_L^{(n)}\right)} - e^{-\theta_n \overline{C}_n}\right] + \sum_{L=1}^{K_{\text{bs}}} \zeta_{\mathbf{H},L} \left(\sum_{n=1}^{K_{\text{mu}}} \mathcal{P}_L^{(n)} - \mathcal{P}_L\right) + \sum_{L=0}^{K_{\text{bs}}} \phi_L L \tag{42}$$

under $\sum_{L=1}^{K_{\text{bs}}} \phi_L = 1$, where $\lambda_n \geq 0$ for $n = 1, 2, \ldots, K_{\text{mu}}$ are the Lagrangian multipliers associated with Eq. (34), which are constants over all fading states, and $\boldsymbol{\lambda} \triangleq (\lambda_1, \lambda_2, \ldots, \lambda_{K_{\text{mu}}})$; $\{\zeta_{\mathbf{H},L}\}_{L=1}^{K_{\text{bs}}}$ are the Lagrangian multipliers associated with Eq. (33), which vary with $\mathbf{H}$ and $L$, and $\boldsymbol{\zeta}_{\mathbf{H},L} \triangleq (\zeta_{\mathbf{H},1}, \zeta_{\mathbf{H},2}, \ldots, \zeta_{\mathbf{H},K_{\text{bs}}})$.

Problem $\boldsymbol{A3}$'s Lagrangian dual function [12], [14], denoted by $\widetilde{\mathcal{J}}_{A3}(\boldsymbol{\lambda}, \boldsymbol{\zeta}_{\mathbf{H}})$, is determined by

$$\begin{aligned} \widetilde{\mathcal{J}}_{A3}(\boldsymbol{\lambda}, \boldsymbol{\zeta}_{\mathbf{H}}) &\triangleq \min_{(\boldsymbol{\phi}, \boldsymbol{\mathcal{P}})} \left\{\mathcal{J}_{A3}(\boldsymbol{\phi}, \boldsymbol{\mathcal{P}}; \boldsymbol{\lambda}, \boldsymbol{\zeta}_{\mathbf{H}})\right\} \\ &= \mathbb{E}_{\mathbf{H}}\left\{\min_{\boldsymbol{\phi}, \boldsymbol{\mathcal{P}}}\left\{J_{A3}(\boldsymbol{\phi}, \boldsymbol{\mathcal{P}}; \boldsymbol{\lambda}, \boldsymbol{\zeta}_{\mathbf{H}})\right\}\right\}. \end{aligned} \tag{43}$$

We denote the minimizer pair in Eq. (43) by $(\widetilde{\boldsymbol{\phi}}, \widetilde{\boldsymbol{\mathcal{P}}})$. Then, we can derive

$$\widetilde{\boldsymbol{\phi}} = \arg\min_{\boldsymbol{\phi}: \sum_{L=1}^{K_{\text{bs}}} \phi_L = 1} \left\{\sum_{L=1}^{K_{\text{bs}}} \phi_L \left(L + \sum_{n=1}^{K_{\text{mu}}} \lambda_n e^{-\theta_n R^{(n)}\left(\Omega_L, \widetilde{\mathcal{P}}_L^{(n)}\right)}\right)\right\}, \quad \forall \mathbf{H}, \tag{44}$$

where Eq. (44) holds by plugging Eq. (42) into Eq. (43) and removing the terms independent of $\boldsymbol{\phi}$. Solving Eq. (44), we

obtain

$$\widetilde{\phi}_L=1,\quad \text{if } L=\arg\min_{1\leq\ell\leq K_{\rm bs}}\left\{\ell+\sum_{n=1}^{K_{\rm mu}}\lambda_n e^{-\theta_n R^{(n)}\left(\Omega_\ell,\widetilde{\mathcal{P}}_\ell^{(n)}\right)}\right\}; \tag{45}$$

otherwise, $\widetilde{\phi}_L=0$. Following the above derivations, $\widetilde{\boldsymbol{\mathcal{P}}}_L$ needs to minimize the function $J_{A3}(\boldsymbol{\phi},\boldsymbol{\mathcal{P}};\boldsymbol{\lambda},\boldsymbol{\zeta}_{\mathbf{H}})$ given $\phi_L=1$ and $\phi_j=0$ for all $j\neq L$. Then, we define a set of functions $J_{A3,L}(\boldsymbol{\mathcal{P}};\boldsymbol{\lambda},\boldsymbol{\zeta}_{\mathbf{H},L})$ for $L=1,2,\ldots,K_{\rm bs}$, where $J_{A3,L}(\boldsymbol{\mathcal{P}};\boldsymbol{\lambda},\boldsymbol{\zeta}_{\mathbf{H},L}) \triangleq J_{A3}(\boldsymbol{\phi},\boldsymbol{\mathcal{P}};\boldsymbol{\lambda},\boldsymbol{\zeta}_{\mathbf{H},L})|_{\phi_L=1;\phi_j=0,j\neq L}$. Taking the derivative of $J_{A3,L}(\boldsymbol{\mathcal{P}};\boldsymbol{\lambda},\boldsymbol{\zeta}_{\mathbf{H},L})$ w.r.t. $\mathcal{P}_L^{(n)}$ and letting the derivative equal to zero, we get

$$\zeta_{\mathbf{H},L}-BT\lambda_n\theta_n\mu_L^{(n)}e^{-\theta_n R^{(n)}\left(\Omega_L,\mathcal{P}_L^{(n)}\right)}=0 \tag{46}$$

for all $n$, $L$, and $\mathbf{H}$, where $\mu_L^{(n)}=dR^{(n)}\big(\Omega_L,\mathcal{P}_L^{(n)}\big)/d\mathcal{P}_L^{(n)}$ as given in Eq. (35). Plugging Eq. (36)-(b) into Eq. (46) and solving for the optimal $\mu_L^{(n)}$ under the boundary condition of $\mu_L^{(n)}\geq 1/\varepsilon_{L,1}^{(n)}$, we obtain Eq. (39) with $i=i^*$. Since Eq. (36) is obtained under the condition of $\mu_L^{(n)}\in\big[1/\varepsilon_{L,i}^{(n)},1/\varepsilon_{L,i+1}^{(n)}\big)$, the variable $i^*$ in Eq. (39) must satisfy the condition of $\mu_L^{(n)}\in\big[1/\varepsilon_{L,i^*}^{(n)},1/\varepsilon_{L,i^*+1}^{(n)}\big)$, as shown in Eq. (38). Moreover, we can show that $J_{A3,L}(\boldsymbol{\mathcal{P}};\boldsymbol{\lambda},\boldsymbol{\zeta}_{\mathbf{H},L})$ is a strictly convex function, and thus $i^*$ for Eq. (39) is unique.

The Lagrangian duality principle [12] shows that $\widetilde{\mathcal{J}}(\boldsymbol{\lambda},\boldsymbol{\zeta}_{\mathbf{H}})=\mathcal{J}(\widetilde{\boldsymbol{\phi}},\widetilde{\boldsymbol{\mathcal{P}}};\boldsymbol{\lambda},\boldsymbol{\zeta}_{\mathbf{H}})$ is concave over $\boldsymbol{\lambda}$ and $\boldsymbol{\zeta}_{\mathbf{H}}$. Moreover, the original problem (also called the primal problem) $\boldsymbol{A3}$'s dual problem is defined by $\max_{(\boldsymbol{\lambda},\boldsymbol{\zeta}_{\mathbf{H}})}\big\{\widetilde{\mathcal{J}}(\boldsymbol{\lambda},\boldsymbol{\zeta}_{\mathbf{H}})\big\}$. We denote the optimal objective of $\boldsymbol{A3}$ by $\overline{L}^*$. The equation $\overline{L}^*\geq\max_{(\boldsymbol{\lambda},\boldsymbol{\zeta}_{\mathbf{H}})}\big\{\widetilde{\mathcal{J}}(\boldsymbol{\lambda},\boldsymbol{\zeta}_{\mathbf{H}})\big\}$ always holds [12], where the difference between $\overline{L}^*$ and $\max_{(\boldsymbol{\lambda},\boldsymbol{\zeta}_{\mathbf{H}})}\big\{\widetilde{\mathcal{J}}(\boldsymbol{\lambda},\boldsymbol{\zeta}_{\mathbf{H}})\big\}$ is known as the duality gap [12]. We can further show that $\widetilde{\mathcal{J}}(\boldsymbol{\lambda},\boldsymbol{\zeta}_{\mathbf{H}})$ is differentiable w.r.t. $\boldsymbol{\lambda}$ and $\boldsymbol{\zeta}_{\mathbf{H}}$. Then, based on the Lagrangian duality principle [12], we can obtain

$$\begin{cases}\frac{\partial\widetilde{\mathcal{J}}(\boldsymbol{\lambda},\boldsymbol{\zeta}_{\mathbf{H}})}{\partial\lambda_n} = \mathbb{E}_{\mathbf{H}}\Big\{\sum_{L=0}^{K_{\rm bs}}\widetilde{\phi}_L e^{-\theta_n R^{(n)}\left(\Omega_L,\widetilde{\mathcal{P}}_L^{(n)}\right)} \\ \qquad\qquad\qquad -e^{-\theta_n\overline{C}_n}\Big\}, \quad \forall n; \\ \frac{\partial\widetilde{\mathcal{J}}(\boldsymbol{\lambda},\boldsymbol{\zeta}_{\mathbf{H}})}{\partial\zeta_{\mathbf{H},L}} = \Big(\sum_{n=1}^{K_{\rm mu}}\widetilde{\mathcal{P}}_L^{(n)}-\mathcal{P}_L\Big)g(\mathbf{H})d\mathbf{H}, \quad \forall L,\, \mathbf{H}\end{cases} \tag{47}$$

where $g(\mathbf{H})$ is the probability density function of $\mathbf{H}$ and $d\mathbf{H}$ denotes the integration variable. As a result, the maximizer $\zeta_{\mathbf{H},L}^*$ must be selected such that $\sum_{n=1}^{K_{\rm mu}}\widetilde{\mathcal{P}}_L^{(n)}-\mathcal{P}_L=0$. Such a $\zeta_{\mathbf{H},L}^*$ exists because $\zeta_{\mathbf{H},L}\to 0$ and $\zeta_{\mathbf{H},L}\to\infty$ leads to $\widetilde{\mathcal{P}}_L^{(n)}\to\infty$ and $\widetilde{\mathcal{P}}_L^{(n)}\to 0$, respectively, as indicated by Eqs. (36)-(a) and (39).

Having obtained $\boldsymbol{\zeta}_{\mathbf{H}}^*$, we next focus on the optimal $\boldsymbol{\lambda}^*$ to maximize $\widetilde{\mathcal{J}}(\boldsymbol{\lambda},\boldsymbol{\zeta}_{\mathbf{H}}^*)$. Due to the concavity of $\widetilde{\mathcal{J}}_{A3}(\boldsymbol{\lambda},\boldsymbol{\zeta}_{\mathbf{H}})$, $\partial\widetilde{\mathcal{J}}(\boldsymbol{\lambda},\boldsymbol{\zeta}_{\mathbf{H}}^*)/\partial\lambda_n$ is a decreasing function of $\lambda_n$. Also, we can readily show that $\partial\widetilde{\mathcal{J}}(\boldsymbol{\lambda},\boldsymbol{\zeta}_{\mathbf{H}}^*)/\partial\lambda_n|_{\lambda_n=0}>0$. Then, if there does not exist $\boldsymbol{\lambda}$ such that $\partial\widetilde{\mathcal{J}}(\boldsymbol{\lambda},\boldsymbol{\zeta}_{\mathbf{H}}^*)/\partial\lambda_n=0$ for all $n$, we have $\lambda_n^*\to\infty$ for some $n$th user and $\partial\widetilde{\mathcal{J}}(\boldsymbol{\lambda},\boldsymbol{\zeta}_{\mathbf{H}}^*)/\partial\lambda_n>0$ always holds. For this case, we get $\overline{L}^*\geq\widetilde{\mathcal{J}}(\boldsymbol{\lambda}^*,\boldsymbol{\zeta}_{\mathbf{H}}^*)\to\infty$, implying that there is no feasible solution for $\boldsymbol{A3}$.

In contrast, if there exists $\boldsymbol{\lambda}^*$ such that $\partial\widetilde{\mathcal{J}}(\boldsymbol{\lambda}^*,\boldsymbol{\zeta}_{\mathbf{H}}^*)/\partial\lambda_n=0$ for all $n$, the pair of $(\boldsymbol{\lambda}^*,\boldsymbol{\zeta}_{\mathbf{H}}^*)$ is the optimal solution to the dual problem given by Eq. (47). Plugging $\partial\widetilde{\mathcal{J}}(\boldsymbol{\lambda}^*,\boldsymbol{\zeta}_{\mathbf{H}}^*)/\partial\lambda_n=0$ into Eq. (47), we can see that the effective-capacity constraint required by Eq. (34) is satisfied for every user. Also note that the optimum for $\max_{(\boldsymbol{\lambda},\boldsymbol{\zeta}_{\mathbf{H}})}\big\{\widetilde{\mathcal{J}}(\boldsymbol{\lambda},\boldsymbol{\zeta}_{\mathbf{H}})\big\}$ is achieved by using $(\widetilde{\boldsymbol{\mathcal{P}}},\widetilde{\boldsymbol{\phi}})$ under $\boldsymbol{\lambda}^*$ and $\boldsymbol{\zeta}_{\mathbf{H}}^*$, which satisfy the constraints imposed by Eqs. (32)-(33). Therefore, all constraints for problem $\boldsymbol{A3}$ are satisfied, implying that this policy is feasible to problem $\boldsymbol{A3}$. As a result, $\overline{L}^*=\max_{(\boldsymbol{\lambda},\boldsymbol{\zeta}_{\mathbf{H}})}\big\{\widetilde{\mathcal{J}}(\boldsymbol{\lambda},\boldsymbol{\zeta}_{\mathbf{H}})\big\}$ holds with zero duality gap. Thus, this policy is the optimal solution to $\boldsymbol{A3}$. Then, setting $\boldsymbol{\mathcal{P}}^*=\widetilde{\boldsymbol{\mathcal{P}}}$ and $\boldsymbol{\phi}^*=\widetilde{\boldsymbol{\phi}}$ with $\boldsymbol{\lambda}^*$ and $\boldsymbol{\zeta}_{\mathbf{H}}^*$ in Eq. (45), we obtain Eqs. (40)-(41). Further plugging Eq. (39) into Eq. (36)-(a), we prove that Eq. (37) holds. Finally, comparing $\partial\widetilde{\mathcal{J}}(\boldsymbol{\lambda}^*,\boldsymbol{\zeta}_{\mathbf{H}}^*)/\partial\lambda_n=0$ with Eq. (47), we show that the equality of Eq. (34) holds, which completes the proof of Theorem 3. ■

Note that there are no closed-form solutions for the optimal Lagrangian multipliers $\boldsymbol{\zeta}_{\mathbf{H},L}^*$ and $\boldsymbol{\lambda}^*$. However, we can determine the values of $\boldsymbol{\zeta}_{\mathbf{H},L}^*$ and $\boldsymbol{\lambda}^*$ by using the numerical searching method similar to the approach given in Appendix B. The detailed searching techniques are omitted due to lack of space, but are provided on-line in the technical-report version [19] of this paper.

### D. The TDMA Based BS-Selection Scheme

We next propose the TDMA based Q0S-aware BS-selection scheme for the comparative analyses. In the TDMA based BS-selection, we also apply the priority BS-selection algorithm given by Fig. 4 when transmission mode $L$ is specified. For transmission mode $L$, we further divide each time frame into $K_{\rm mu}$ time slots for data transmissions to $K_{\rm mu}$ users, respectively. The $n$th user's time-slot length is set equal to $T\times t_{L,n}$ for $n=1,2,\ldots,K_{\rm mu}$, where $t_{L,n}$ is the normalized time-slot length. Moreover, we still use the probabilistic transmission strategy across different transmission modes, where the probability of using transmission mode $L$ to transmit data is equal to $\phi_L$. Then, we derive the TDMA and probabilistic transmission policies through solving the following optimization problem $\boldsymbol{A4}$.

$$\boldsymbol{A4}:\ \min_{(\boldsymbol{t}(\mathbf{H}),\boldsymbol{\phi}(\mathbf{H}))}\{\overline{L}\}=\min_{(\boldsymbol{t}(\mathbf{H}),\boldsymbol{\phi}(\mathbf{H}))}\left\{\mathbb{E}_{\mathbf{H}}\left\{\sum_{L=0}^{K_{\rm bs}}L\phi_L(\mathbf{H})\right\}\right\}$$

$$\text{s.t.: } 1).\ \sum_{L=0}^{K_{\rm bs}}\phi_L(\mathbf{H})=1,\ \forall\mathbf{H}, \tag{48}$$

$$2).\ \sum_{n=1}^{K_{\rm mu}}t_{L,n}(\mathbf{H})=1,\ \forall\mathbf{H},\ L=1,2,\ldots,K_{\rm bs}, \tag{49}$$

$$3).\ \mathbb{E}_{\mathbf{H}}\left\{\sum_{L=0}^{K_{\rm bs}}\phi_L(\mathbf{H})e^{-\theta_n t_{L,n}(\mathbf{H})R^{(n)}(\Omega_L,\mathcal{P}_L)}\right\} \leq e^{-\theta_n\overline{C}_n},\quad \forall n, \tag{50}$$

where $\boldsymbol{\phi}(\mathbf{H})$ and $\boldsymbol{t}(\mathbf{H})$ are the probabilistic transmission vector and the time-division policy. In particular, in problem $\boldsymbol{A4}$ we have $\boldsymbol{\phi}(\mathbf{H})\triangleq(\phi_0(\mathbf{H}),\phi_1(\mathbf{H}),\ldots,\phi_{K_{\rm mu}}(\mathbf{H}))$ and $\boldsymbol{t}(\mathbf{H})\triangleq(\boldsymbol{t}_1(\mathbf{H}),\boldsymbol{t}_2(\mathbf{H}),\ldots,\boldsymbol{t}_{K_{\rm bs}}(\mathbf{H}))$ with $\boldsymbol{t}_L\triangleq$ $(t_{L,1}(\mathbf{H}), t_{L,2}(\mathbf{H}), \ldots, t_{L,K_{\text{bs}}}(\mathbf{H}))$. We use $\boldsymbol{\phi}(\mathbf{H})$ and $\boldsymbol{t}(\mathbf{H})$ to denote the probabilistic transmission vector and the time-division policy to address that they are function of $\mathbf{H}$. When the context is clear, we drop the symbol $(\mathbf{H})$ in problem $\boldsymbol{A4}$ to simplify notations. We call the optimal solution of problem $\boldsymbol{A4}$ as the PBS-TDMA-PT scheme.

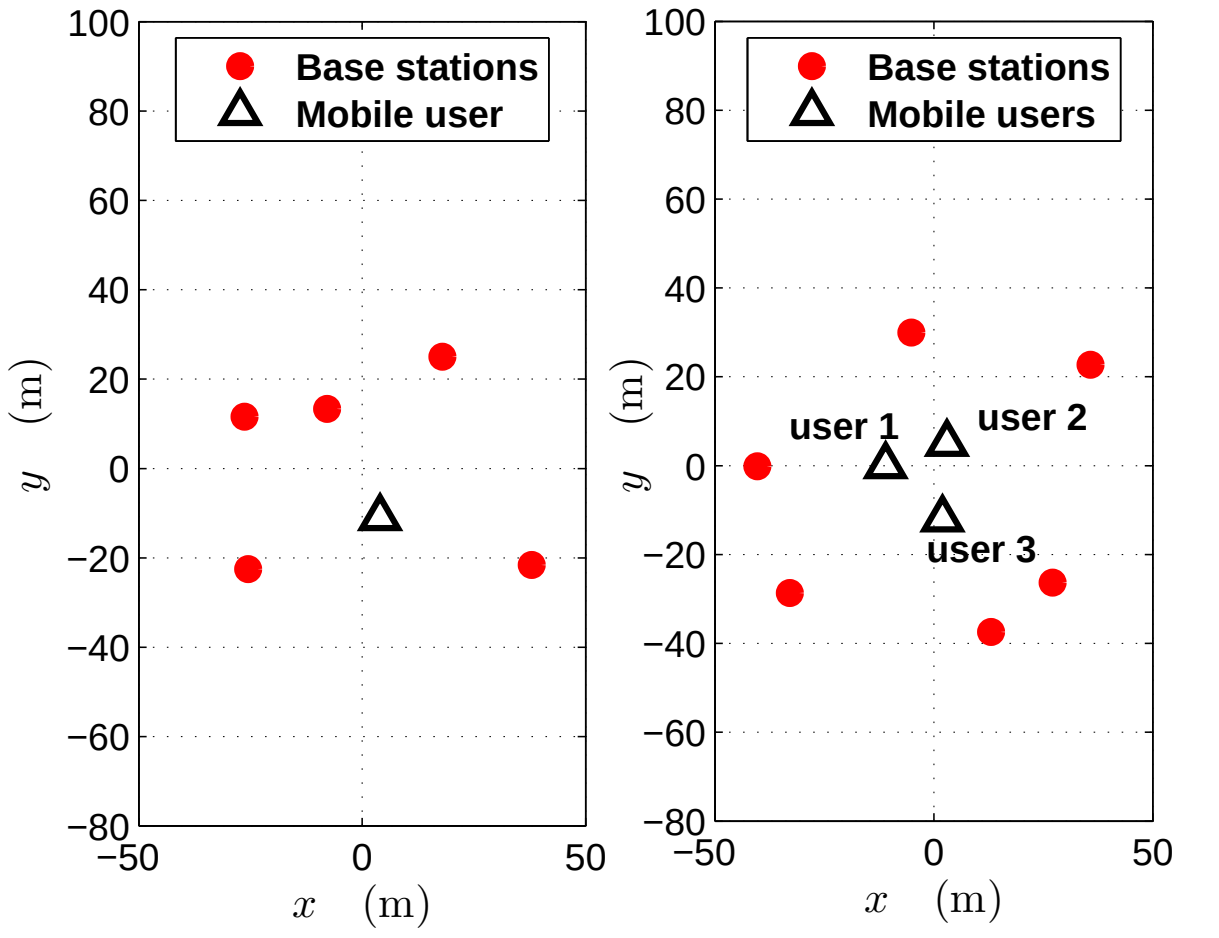

(a) (b)

Fig. 5. The deployment of BS's and the positions of mobile users. (a) Single-user case: $K_{\text{bs}} = 5$ BS's, whose coordinates are $(37.96, -21.56)$, $(-7.83, 13.33)$, $(25.50, -22.49)$, $(17.98, 25.00)$, and $(-26.34, 11.62)$; the mobile station's coordinates are $(4, -11)$. (b) Multi-user case: $K_{\text{bs}} = 6$, whose coordinates are $(35.77, 22.69)$, $(13.06, -37.45)$, $(27.15, -26.33)$, $(-40.28, -0.14)$, $(-32.86, -28.65)$, and $(-5.10, 29.98)$; $K_{\text{mu}} = 3$, whose coordinates are $(-11, 0)$, $(3, 5)$, and $(2, -12)$.

*Theorem 4:* Problem $\boldsymbol{A4}$'s optimal solution pair $(\boldsymbol{t}^*, \boldsymbol{\phi}^*)$, if existing, is determined by

$$t^*_{L,n} = \left[\frac{1}{\theta_n R^{(n)}(\Omega_L, \mathcal{P}_L)} \log\left(\frac{\lambda^*_n \theta_n R^{(n)}(\Omega_L, \mathcal{P}_L)}{\delta^*_{\mathbf{H},L}}\right)\right]^+, \quad (51)$$

for all $L$, $n$, and $\mathbf{H}$, and

$$\phi^*_L = \begin{cases} 1, & \text{if } L = \arg\min\limits_{\ell=0,1,\ldots,K_{\text{bs}}} \Big\{\ell \\ & \quad + \sum_{n=1}^{K_{\text{mu}}} \lambda^*_n e^{-\theta_n t^*_{L,n} R^{(n)}(\Omega_\ell, \mathcal{P}_\ell)}\Big\}; \\ 0, & \text{otherwise} \end{cases} \quad (52)$$

for all $L$ and $\mathbf{H}$, where $\delta^*_{\mathbf{H},L}$ under given $\{\lambda^*_n\}_{n=1}^{K_{\text{mu}}}$ is determined by satisfying $\sum_{n=1}^{K_{\text{mu}}} t^*_{L,n} = 1$, and $\{\lambda^*_n\}_{n=1}^{K_{\text{mu}}}$ needs to be selected such that the equality of Eq. (50) holds.

*Proof:* The detailed proof of Theorem 4 is omitted due to lack of space, but is provided on-line in the technical-report version [19] of this paper. ■

## VI. Simulation Evaluations

We use simulations to evaluate the performances of our proposed QoS-aware BS-selection schemes for distributed MIMO links. In particular, for the single-user case, the BS's deployment and the mobile user's position are shown in Fig. 5(a), where $K_{\text{bs}} = 5$. For the multi-user case, the BS's deployment and the mobile users' positions are given by Fig. 5(b), where $K_{\text{bs}} = 6$ and $K_{\text{mu}} = 3$. Moreover, we set $T = 10$ ms and $B = 10^5$ Hz throughout the simulations. In the simulations, we focus on the case that the positions of users do not change. This can be mapped to the scenarios where users move slowly (e.g., pedestrians) or do not move. In such scenarios, the users' statistical channel conditions, depending on the distances between users and distributed BS's, often change very slightly in tens of seconds, during which many (thousands of) packets will be transmitted before the users' statistical channel conditions vary significantly. Then, our developed statistical delay-QoS aware BS-selection schemes, which are based on the current statistical channel information, can effectively guarantee the corresponding delay QoS during this period. For simulations of scenarios that the users move very quickly, the delay-QoS-aware BS-selection schemes highly depend on the users' moving speeds and trajectories. These requirements are different from the requirements on typical simulation environments [23] for transmission error rate evaluations in cellular networks, where only the distribution information of users' positions are needed. As a result, the delay-QoS-aware BS-selection schemes to be re-designed, which will be studied in our future work.

In the simulations, we employ the following average power degradation propagation model [15]. For the given reference distance $d_{\text{ref}}$, if the transmission distance, denoted by $d$, is smaller than or equal to $d_{\text{ref}}$, the free-space propagation model is used; if $d > d_{\text{ref}}$, the power degradation is proportional to $(d/d_{\text{ref}})^\eta$, where $\eta$ is the path loss exponent and typically varies from 2 to 6 indoor environments without LOS [15]. Accordingly, the variance $\overline{h}_{n,m}$ of $H_{n,m}$'s elements can be determined by

$$\overline{h}_{n,m} = \begin{cases} G d_{n,m}^{-2}, & \text{if } d_{n,m} \in (0, d_{\text{ref}}]; \\ G\left(\frac{d_{\text{ref}}}{d_{n,m}}\right)^\eta, & \text{if } d_{n,m} \in (d_{\text{ref}}, \infty), \end{cases}$$

where $d_{n,m}$ is the distance between the mobile user and the $m$th BS, and $G$ is aggregate power gain generated by the antenna and other factors. In simulations, we set $d_{\text{ref}} = 1$ m [15] and $\eta = 3$. Furthermore, we select $G$ such that $\overline{h}_{n,m} = 0$ dB at $d_{n,m} = 50$ m. Also, we set $\sigma^2_{\text{th}} = 0$ dB for the evaluation of the average interfering range (see Section II-C for its definition).

We also consider the following baseline BS-selection schemes for comparative analyses:

*1). Base-Station Selection with Fixed Cardinality for Single-User Cases*

This scheme does not change the selected BS-subsect's cardinality over all fading states. The fixed cardinality, denoted by $L_{\text{fix}}$, is determined by

$$L_{\text{fix}} = \min\Big\{L \,\Big|\, \mathbb{E}_{\mathbf{H}}\left\{e^{-\theta_1 R(\Omega_L)}\right\} \le \mathbb{E}_{\mathbf{H}}\left\{e^{-\theta_1 \overline{C}_1}\right\}, L \text{ is fixed}\Big\}, \quad (53)$$

where $\Omega_L$ for each $\mathbf{H}$ is determined through the ordered-gain based BS-selection algorithm given by Section IV-A, as also used by our derived OGBS-PT scheme.

*2). Optimal Time-Sharing Based Scheme for Single-User Cases*

The optimal time-sharing (TS) based scheme is the same as our proposed IBS-TS scheme except that it uses exhaustive search to find the optimal $\Omega_L$ for Eq. (13) in each fading state. It is clear that the optimal TS based scheme provides the lower-bound of the average BS usage for our proposed IBS-TS scheme.

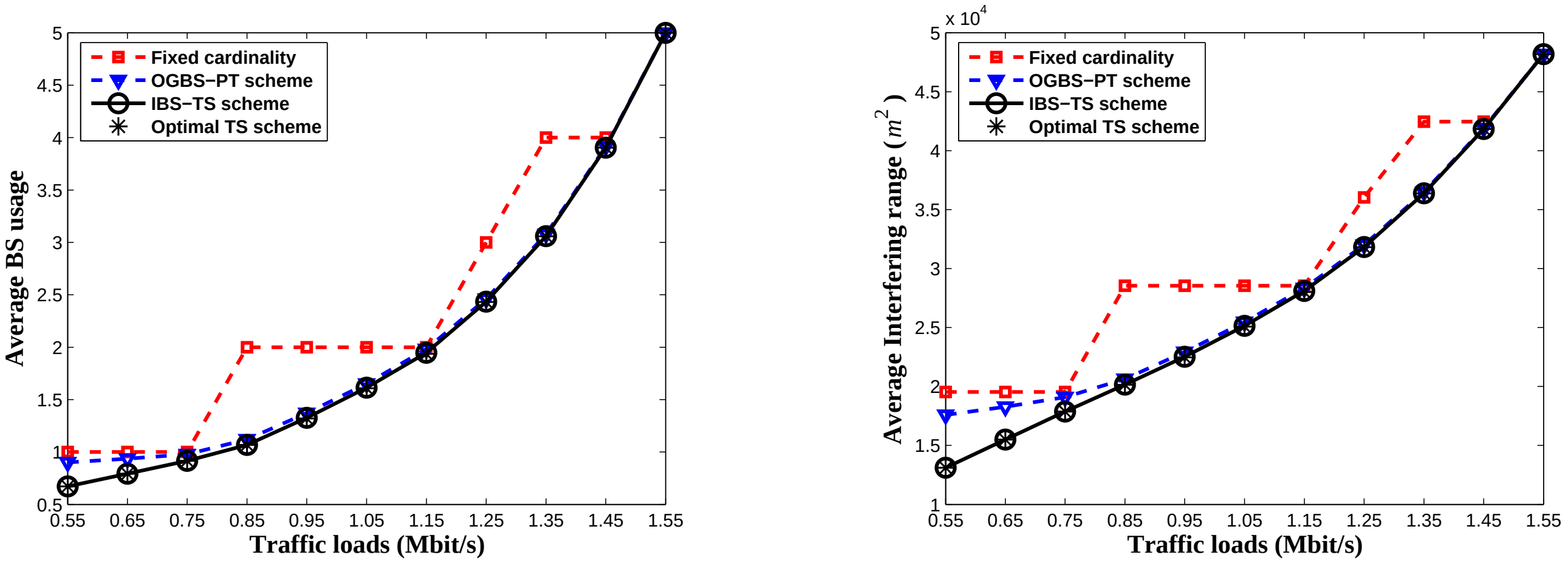

(a) (b)

Fig. 6. Simulations of the single-user case for the specified delay-QoS requirements given by $D_{\rm th}^{(1)} = 50$ ms and $\xi_1 = 10^{-4}$, where $K_{\rm bs} = 5$, each BS has two transmit antennas, the mobile user has two receive antennas, $\mathcal{P}_{\rm ref} = 4$, and $\kappa = 2.4$. (a) Average BS usage versus traffic load. (b) Average interfering range versus traffic load.

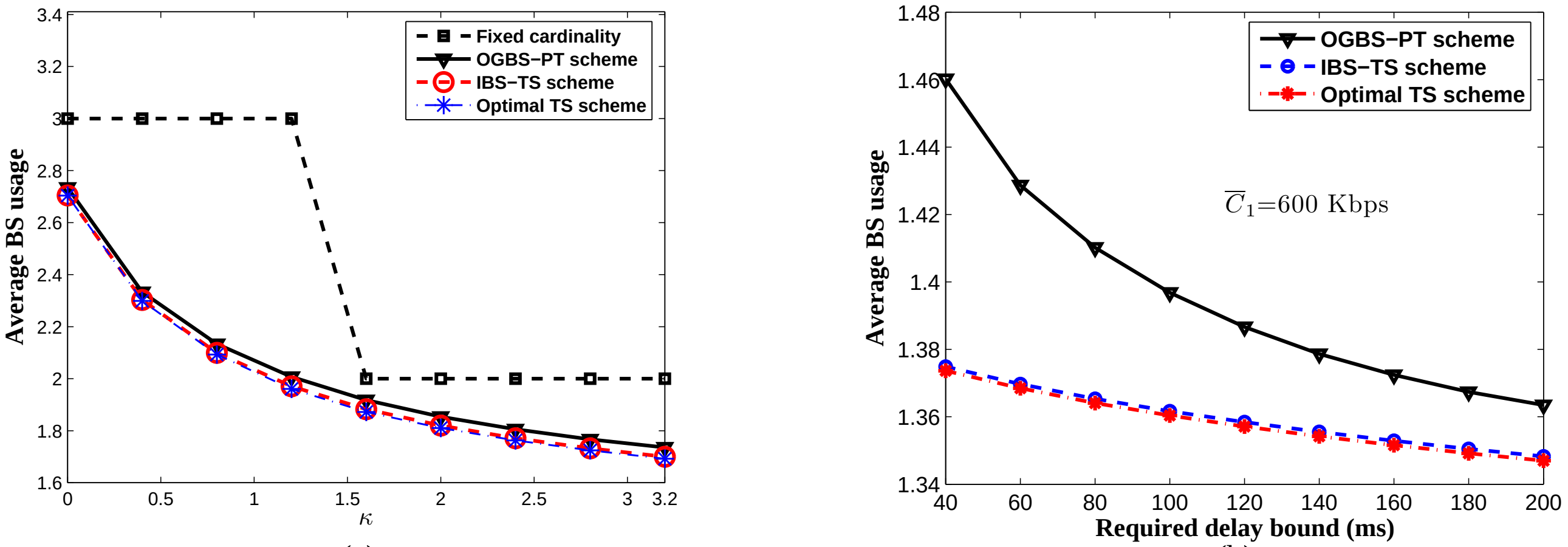

(a) (b)

Fig. 7. Simulation results for single-user case varying the specified delay-QoS requirements and system parameters, where $K_{\rm bs} = 5$, each BS has two transmit antennas, the mobile user has two receive antennas. (a) Average BS usage versus $\kappa$, where $D_{\rm th}^{(1)} = 50$ ms, $\xi_1 = 10^{-4}$, $\overline{C}_1 = 1.1$ Mbps, and $\mathcal{P}_{\rm ref} = 1$. (b) Average BS usage versus delay-bound QoS requirements, where $\xi_1 = 10^{-6}$, $\overline{C}_1 = 600$ Kbps, $\kappa = 0.4$, and $\mathcal{P}_{\rm ref} = 1$.

*3). Semi-Random BS-Selection Scheme for Multi-user Cases*

In this scheme, given the cardinality $L$, the central controller randomly selects $\Omega_L$ with the uniform distribution across all BS's. Furthermore, the cardinality $L$ of the selected BS subset is determined by using the algorithm derived through Theorem 3. We call this scheme as the *semi-random BD-PT* scheme in that the selection of $L$ is not in a random manner. We evaluate the semi-random BD-PT scheme to provide an upper-bound of the average BS-usage for our derived priority BS-selection algorithm given in Fig. 4. This semi-random strategy can be also applied to the TDMA based scheme, where $\Omega_L$ is randomly selected and $L$ is determined by using the algorithm obtained through Theorem 4.

Figures 6(a) and 6(b) plot the average BS usage and the average interfering range, respectively, versus the incoming traffic loads for the single-user case. As shown in Fig. 6(a), our proposed IBS-TS and OGBS-PT schemes both effectively decrease the average BS usage and the interfering range as compared to the fixed BS-selection scheme. This is expected because our proposed BS-selections can adaptively adjust BS selection in each fading state based on the CSI, traffic loads, and the QoS requirements. In contrast, the resulted average BS usage and the interfering range by applying the fixed BS-selection scheme cannot smoothly vary with traffic loads, which may cause unnecessary BS usage with more power consumption and thus larger interferences to the entire wireless networks. We also observe from Fig. 6 that the IBS-TS scheme needs less BS usage than the OGBS-PT scheme to support the incoming traffic loads under the specified QoS requirements and therefore generates lower interferences accordingly, which verifies our discussions in Sections IV-A and IV-B. However, we can see that the performance differences between the IBS-TS and OGBS-PT schemes are slight, especially when the incoming traffic loads are relatively high. In addition, we can observe from Fig. 6 that the BS-usage performance and the interfering range of our proposed IBS-TS scheme are almost the same as compared to the optimal TS scheme, which verifies

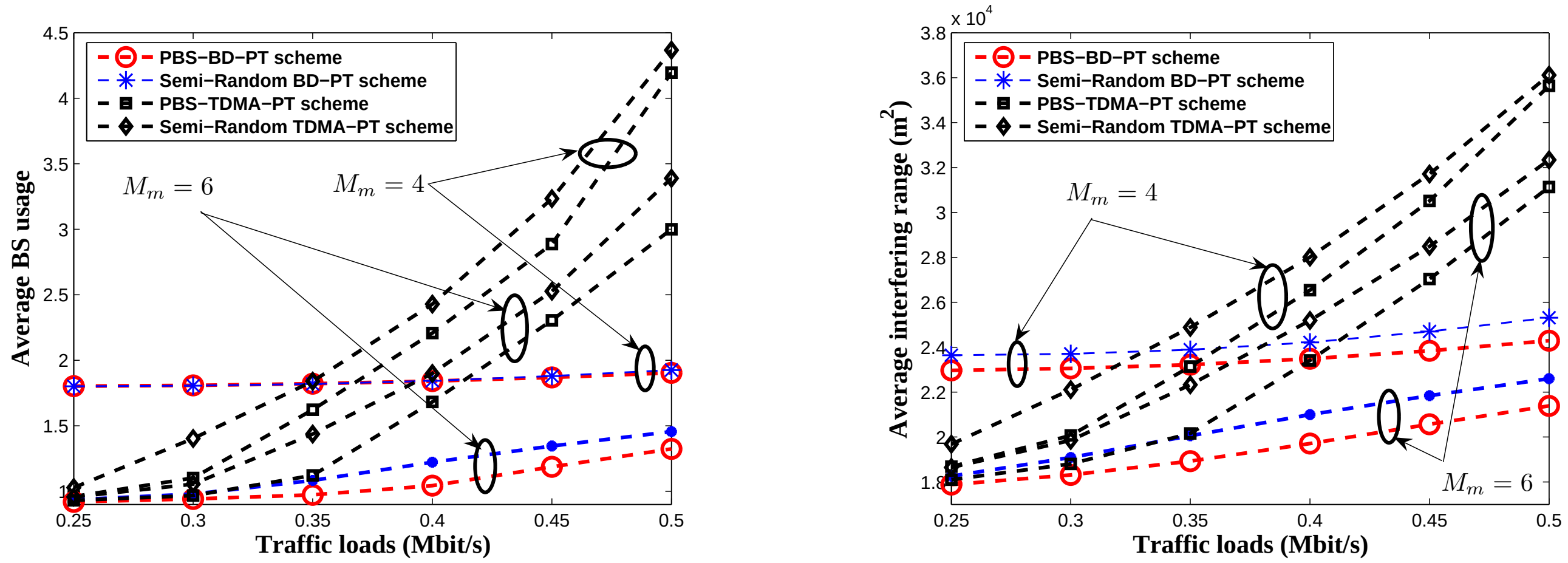

(a) (b)

Fig. 8. Simulation results of the multi-user case for the specified delay-QoS requirements given by $\xi_n = 10^{-4}$ for all $n$ and $D_{\rm th}^{(1)} = D_{\rm th}^{(2)} = 50$ ms and $D_{\rm th}^{(3)} = 40$ ms, where $K_{\rm bs} = 6$, $K_{\rm mu} = 3$, $N_n = 2$ for all $n = 1, 2, \ldots, K_{\rm mu}$, $\mathcal{P}_{\rm ref} = 4$, $\kappa = 1.2$, and $M_m$ is the same for all users. (a) Average BS usage versus traffic load. (b) Average interfering range versus traffic load.

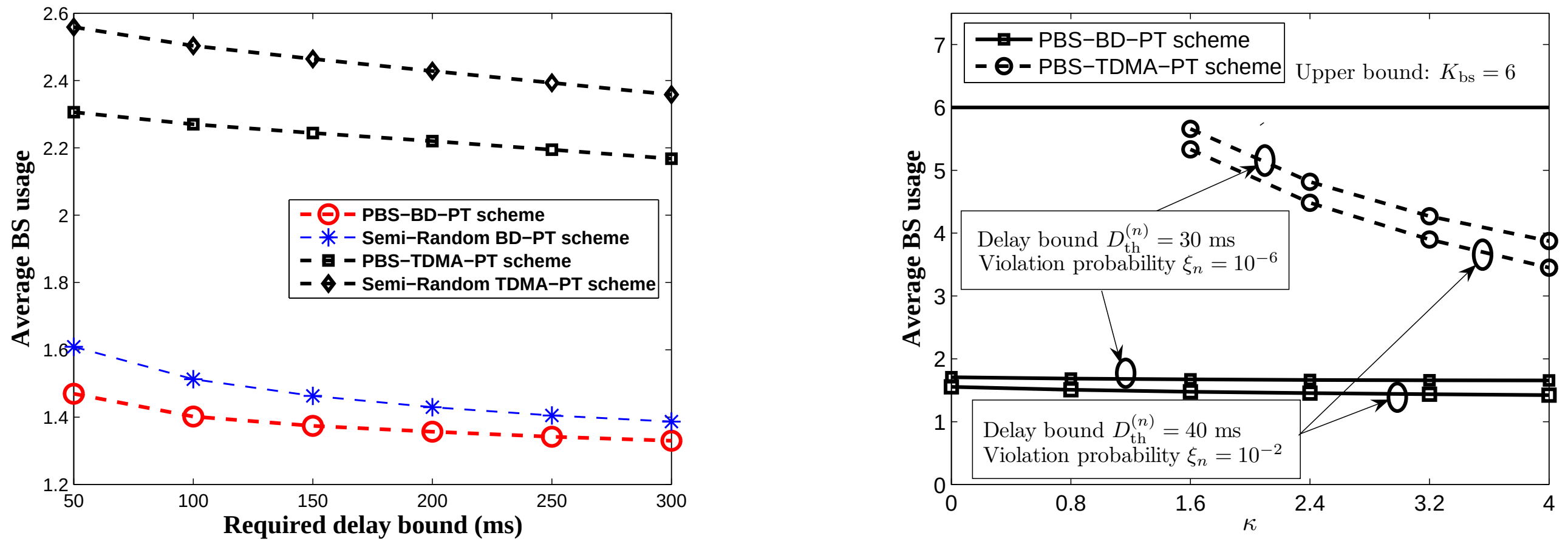

(a) (b)

Fig. 9. Simulation results for impact of QoS requirement on the multi-user case, where $K_{\rm bs} = 6$, $K_{\rm mu} = 3$, and $N_n = 2$ for all $n = 1, 2, \ldots, K_{\rm mu}$ and $M_m = 6$ for all $m = 1, 2, \ldots, K_{\rm bs}$; $\{\xi_n\}_{n=1}^{K_{\rm mu}}$ are equal; $\{D_{\rm th}^{(n)}\}_{n=1}^{K_{\rm mu}}$ are equal; $\{\overline{C}_n\}_{n=1}^{K_{\rm mu}}$ are equal. (a) Average BS usage versus delay-bound QoS requirements, where $\overline{C} = 300$ Kbps, $\xi_n = 10^{-4}$, $\kappa = 0.1$, and $\mathcal{P}_{\rm ref} = 1$. (b) Average BS usage versus $\kappa$, where $\overline{C} = 600$ Kbps and $\mathcal{P}_{\rm ref} = 4$.

the effectiveness of our proposed IBS-TS scheme.

Figure 7(a) depicts the dynamics of the average BS usage and the interfering range as functions of $\kappa$, where $\kappa$ is the power increasing rate against the number of selected BS as defined in Eq. (3). When $\kappa$ gets larger, the power budget used for distributed MIMO transmission in each fading state increases. As a result, the average BS usage is reduced correspondingly for all schemes, as illustrated in Fig. 7(a). Again, Fig. 7(a) demonstrates the inflexibility of the fixed BS-selection, and the significant BS-usage reduction and the interfering-range reduction as a result of using our proposed IBS-TS and OGBS-PT schemes.

Figure 7(b) illustrates the impact of delay-QoS requirements on the average BS-usage among different BS-selection algorithms. As shown in Figures 7(b), the average BS-usage decreases as the required delay-bound increases, which is expected because of the following reasons. When the delay QoS constraint becomes looser, suggesting a larger delay bound, the central controller does not have to use more BS's to improve the transmission rate when the wireless channel experiences deep fading. Correspondingly, the efficiency of using the distributed BS's can be improved, thus reducing the average BS usage. The above observations verify that our proposed BS-selection algorithms can effectively adapt to the variation of delay QoS requirements. Also, from Figure 7(b) we can observe again that the IBS-TS results in less BS usage than the OGBS-PT scheme, and the IBS-TS scheme achieves almost the same performance as the optimal TS scheme.

Figure 8 compares the average BS usage and interfering range among our proposed BS-selection scheme for multi-user case under various system parameters. Fig. 8 shows that as the traffic loads increase, all scheme's average BS usages and interfering ranges increase to satisfy the specified QoS requirements for the incoming traffic loads. Moreover, the PBS-TDMA-PT scheme's BS usage increases much more rapidly than the PBS-BD-PT scheme. As shown in Fig. 8(a), the PBS-BD-PT scheme results in less average BS usage than the PBS-TDMA-PT scheme when the traffic load is high. The superiority of the PBS-BD-PT over the PBS-TDMA-PT scheme gradually vanishes as the traffic loads become low. Furthermore, we can observe that given $M_m = 6$, the PBS-BD-PT scheme achieves lower BS usage than the PBS-TDMA-PT scheme under various traffic loads. This is expected since it has been demonstrated that the BD technique usually outperforms the TDMA technique in achieving high throughput. However, for $M_m = 4$, the PBS-TDMA-PT scheme leads to less BS usage when the traffic load is relatively low. The above result is caused by the following reasons. In order to support simultaneous transmissions across multiple users by using the BD technique, a necessary condition is that the total number of transmit antennas over all distributed BS's can provide enough freedoms for spatial multiplexing. Thus, if the number $M_m$ of transmit antenna per BS is smaller, more BS's have to involve the transmission to gain enough spatial-multiplexing freedoms, even when the traffic load is low. In contrast, when the traffic load is very low, TDMA does not have such requirement on spatial multiplexing because it achieves cross-interference-free transmissions through time-slot division. The above observations will motive us to develop hybrid TDMA-BD BS-selection schemes in future studies.

Figure 8(a) further illustrates that our proposed PBS-BD-PT and PBS-TDMA-PT schemes in general yields less BS usages than the semi-random BD-PT and the semi-random TDMA-PT schemes, respectively. Note that when $M_m = 4$, the BS-usages of the semi-random BD-PT scheme and our proposed PBS-BD-PT scheme are the same, which also increase very slowly as the traffic loads increase. This is because for small number of transmit antennas, more BS's are needed to support the space-multiplexing, as aforementioned. When more BS's are used, the achievable rate for each dimension/user often exceed the traffic load for the corresponding user when the traffic load is low. In such a case, even randomly selecting BS can satisfy the specified delay-QoS constraint under the given traffic loads. However, even when the BS usages of the semi-random BD-PT scheme and our proposed PBS-BD-PT scheme are the same, our proposed we can see from that Fig. 8(b) that the PBS-BD-PT scheme results in a smaller interfering range by selecting the BS based CSI.

Figure 9(a) plots the dynamics of the BS usage versus the required delay bound. As depicted in Fig. 9(a), the larger delay bound leads to less BS usage for all schemes, which demonstrates that our proposed schemes can effectively adjust the BS selections to meet diverse QoS requirements across multiple co-existing users. Also, we can see that the semi-random based schemes cause significantly higher BS-usages than our derived PBS based schemes. Fig. 9(b) illustrates our proposed PBS-BD-PT scheme and PBS-TDMA-PT based scheme's performances versus $\kappa$ (see Eq. (3)) under diverse QoS requirements. Similar to the single-user case, the average BS usage deceases as $\kappa$ gets larger. Moreover, as shown in Fig. 9, the lower delay bound and the smaller violation probability threshold, implying more stringent delay-QoS requirements, cause more BS usages and thus larger interfering ranges.

## VII. Conclusions

We proposed the QoS-aware BS-selection schemes for the distributed wireless MIMO downlink to minimize the BS usages and to reduce the interfering range affected by the distributed MIMO system, while satisfying diverse statistical delay-QoS constraints characterized by the delay-bound violation probability and the effective capacity technique. For the single-user scenario, we developed the scheme using the incremental BS selection and time-sharing strategy and also developed the scheme employing the ordered-gain based BS-selection and probabilistic transmission strategy. The former scheme archives better performance, while the latter scheme is easier to implement. For the multi-user scenario, we developed the joint priority BS-selection, block-diagonalization precoding, and probabilistic transmission schemes. We also studied the TDMA based BS-selection scheme for multi-user link. The extensive simulation results show that our proposed schemes can effectively support the incoming traffic loads for the specified QoS requirements and significantly outperform baseline schemes in terms of minimizing the average BS usage and the interfering range.

## Appendix A
## Proof of Theorem 1

*Proof:* We construct the Lagrangian function of $\boldsymbol{A1}$, denoted by $\mathcal{J}_{A1}(\mathcal{L},\lambda)$, given as $\mathcal{J}_{A1}(\mathcal{L},\lambda) = \mathbb{E}_{\mathbf{H}}\{J_{A1}(\mathcal{L},\lambda)\}$ with

$$J_{A1}(\mathcal{L},\lambda) = \mathcal{L} + \lambda\left(e^{-\theta_1\widetilde{R}(\mathcal{L})} - e^{-\theta_1\overline{C}_1}\right), \tag{54}$$

where $\lambda$ is the Lagrangian multiplier associated with the constraint of $\boldsymbol{A1}'$. Then, the optimal $\mathcal{L}^*$ and the optimal Lagrangian multiplier $\lambda^*$ are solutions to the following equations [12]:

$$\begin{cases} 0 \;\in\; \partial_{\mathcal{L}} J_{A1}(\mathcal{L},\lambda), \quad \forall \mathbf{H}; \\ 0 \;=\; \mathbb{E}_{\mathbf{H}}\left\{e^{-\theta_1\widetilde{R}(\mathcal{L})} - e^{-\theta_1\overline{C}_1}\right\}. \end{cases} \tag{55}$$

where $\partial_{\mathcal{L}} J_{A1}(\mathcal{L},\lambda)$ denotes the *subdifferential* [12] of the function $J_{A1}(\mathcal{L},\lambda)$ with respect to $\mathcal{L}$. Note that the subdifferential is defined for nondifferentiable convex functions (e.g., piece-wise linear functions), which is the counterpart concept for the gradient of differentiable convex functions. Based on [12], for a convex function $f(\mathbf{b})$ defined on $\mathbf{b} \in \mathcal{B} \subset \mathbb{R}$, where $\mathbb{R}$ is real-number set and $\mathcal{B}$ is a convex set, an $n \times 1$ real-valued vector $\boldsymbol{\varpi}$ is a *subgradient* of $h(\mathbf{b})$ if $h(\mathbf{b}') \geq h(\mathbf{b}) + \boldsymbol{\varpi}^\tau(\mathbf{b}' - \mathbf{b})$ for all $\mathbf{b}' \in \mathcal{B}$. Accordingly, the collection of subgradients at $\mathbf{b}$ is the *subdifferential* of $h(\mathbf{b})$. For more details and properties of subdifferential, please refer to [12].

Applying the piece-wise linear property and the concavity of $\widetilde{R}(\mathcal{L})$, we derive

$$\partial_{\mathcal{L}} J_{A1}(\mathcal{L},\lambda) = \begin{cases} \left[1-\theta_1\lambda\nu_j e^{-\theta_1\widetilde{R}(m_j)},\, 1-\theta_1\lambda\nu_{j+1} e^{-\theta_1\widetilde{R}(m_j)}\right), \\ \qquad \text{if } \mathcal{L} = m_j,\ j = 0,1,\ldots,\mathcal{K}; \\ \left\{1-\theta_1\lambda\nu_j e^{-\theta_1\widetilde{R}(\mathcal{L})}\right\}, \\ \qquad \text{if } \mathcal{L} \in (m_{j-1}, m_j), j = 1,\ldots,\mathcal{K}. \end{cases} \tag{56}$$

Plugging Eq. (56) into Eq. (55) and solving for the optimal solution, we get Eq. (23). Also, the equality of Eq. (16) needs to hold as required by Eq. (55). which completes the proof of Theorem 1. ■

## APPENDIX B
## THE NUMERICAL-TRACKING METHOD TO DERIVE $\lambda^*$ AND DISCUSSIONS ON ITS COMPLEXITY

As defined in Appendix A, the Lagrangian function under the optimal solution derived in Theorem 1 is given by

$$\mathcal{J}_{A1}(\mathcal{L}^*, \lambda) = \mathbb{E}_{\mathbf{H}}\{\mathcal{L}^*\} + \lambda \mathbb{E}_{\mathbf{H}} \left\{ e^{-\theta_1 \widetilde{R}(\mathcal{L}^*)} - e^{-\theta_1 \overline{C}_1} \right\}.$$

Note that in this equation we remove the superscript $(\cdot)^*$ of $\lambda^*$ because we have not obtained the actual value of $\lambda^*$. Based on convex optimization theory, $\mathcal{J}_{A1}(\mathcal{L}^*, \lambda)$, known as the Lagrangian dual function [12], [14], is concave over $\lambda$ and the derivative $d\mathcal{J}_{A1}(\mathcal{L}^*, \lambda)/d\lambda = \mathbb{E}_{\mathbf{H}}\{e^{-\theta_1 \widetilde{R}(\mathcal{L}^*)} - e^{-\theta_1 \overline{C}_1}\}$. Moreover, if the optimal solution exists, $\mathcal{J}_{A1}(\mathcal{L}^*, \lambda)$ reaches its maximum at $\lambda = \lambda^*$ with $\mathbb{E}_{\mathbf{H}}\{e^{-\theta_1 \widetilde{R}(\mathcal{L}^*)} - e^{-\theta_1 \overline{C}_1}\} = 0$. Therefore, we can use the gradient descent method [14] to search for $\lambda^*$ numerically. The gradient descent method is implemented through the following iterative equation:

$$\lambda := \lambda + \epsilon \mathbb{E}_{\mathbf{H}} \left\{ e^{-\theta_1 \widetilde{R}(\mathcal{L}^*)} - e^{-\theta_1 \overline{C}_1} \right\},$$

where $\epsilon$ is a positive real number much smaller than 1. For the fast tracking of $\lambda^*$, we can use the time average via the first-order autoregressive low-pass filter to estimate and replace the expectation $\mathbb{E}_{\mathbf{H}}\{e^{-\theta_1 \widetilde{R}(\mathcal{L}^*)} - e^{-\theta_1 \overline{C}_1}\}$ for the above gradient descent method. In particular, we generate a large number (more than $10^4$) of channel realizations, representing different fading states, which are indexed by time sequence $k = 1, 2, \ldots$. Denoting the estimate of $\mathbb{E}_{\mathbf{H}}\{e^{-\theta_1 \widetilde{R}(\mathcal{L}^*)} - e^{-\theta_1 \overline{C}_1}\}$ at the $k$-th time by $\mathcal{J}'_\lambda[k]$, we use the following first-order autoregressive low-pass filter to update $\mathcal{J}'_\lambda[k]$:

$$\mathcal{J}'_\lambda[k] = \vartheta \mathcal{J}'_\lambda[k-1] + (1-\vartheta) \left( e^{-\theta_1 \widetilde{R}(\mathcal{L}^*[k])} - e^{-\theta_1 \overline{C}_1} \right) \Big|_{\mathbf{H}=\mathbf{H}[k]},$$

where $\vartheta \in (0, 1)$ is a real number close to 1, $\mathcal{L}^*[k]$ is determined by using Eq. (23) under the current $\lambda$ of the $k$th time frame, and $\mathbf{H}[k]$ denotes the CSI at the $k$th time frame. With appropriately selected $\epsilon$ and $\vartheta$, the above algorithm can converge and oscillate around the optimal $\lambda^*$ within a small range. The above tracking method has been used and validated in many optimization problems for resource allocation over fading channels. By using this algorithm, we can conveniently find the optimal solution without deriving the closed-form expressions for the expectation and $\lambda^*$, which are usually hard to obtain for general fading channel models.

Under the above strategy, the computational complexity for our proposed BS-selection scheme in each fading state is dominated by: 1) the incremental BS-selection algorithm given in Fig. 2 to determine $\Omega_L$; 2) singular value decomposition (SVD) of the channel matrix to calculate the MIMO capacity [5]; 3) the procedures to find the convex hull of $\{(L, \widetilde{R}(L))\}_{i=0}^{K_{\rm bs}}$ for deriving Eq. (19). It is not difficult to see that the complexity of the incremental algorithm is upper-bounded by $O(K_{\rm bs}^2)$; the complexity for SVD is known upper-bounded by $O(K_\Sigma^3)$, where $K_\Sigma \triangleq \sum_{m=1}^{K_{\rm bs}} M_m$; the complexity to identify the convex hull is upper-bounded by $O(K_{\rm bs}\log(K_{\rm bs}))$. Therefore, the total complexity for each fading state is upper-bounded by $O(K_\Sigma^3)$, implying the major computational complexity is caused by the calculation for the MIMO capacity.